\documentclass[12pt]{article}
\usepackage[letterpaper, left=1in, top=1in, right=1in,
bottom=1in,nohead,includefoot, verbose, ignoremp]{geometry}

\usepackage[sort&compress,comma]{natbib}
\usepackage{natbibspacing}
\usepackage{amsfonts}
\usepackage{amsthm}
\usepackage{amsmath,afterpage}
\usepackage{amssymb}
\usepackage{graphicx}
\usepackage{color}
\usepackage{bm}
\usepackage{longtable}
\usepackage{multirow}
\usepackage{tabularx}
\usepackage{dsfont}
\usepackage{setspace}  
\usepackage{url}

\usepackage[compact]{titlesec}
\titlespacing{\section}{0pt}{5pt}{3pt}
\titlespacing{\subsection}{0pt}{5pt}{3pt}
\titlespacing{\subsubsection}{0pt}{4pt}{3pt}

\newcommand{\iid}{\stackrel{\mathrm{iid}}{\sim}}
\newcommand{\refeq}[1]{Eq.~\eqref{#1}} 

\newcommand{\uniqueRate} {\phi}
\newcommand{\tractRate}{\lambda}
\newcommand{\tractAdjRate}{\psi}
\newcommand{\obsVal}{y}
\newcommand{\obsVar}{Y}

\newcommand{\hide}[1]{{}}

\makeindex

\begin{document}
\vskip -1cm
\title{Spatio-Temporal Low Count Processes with Application to Violent Crime Events}

\author{Sivan Aldor-Noiman \and Lawrence D. Brown \and Emily B. Fox \and Robert A. Stine}

\maketitle


\begin{abstract}
\hide{ Violent crimes are a significant source of concern in major metropolitan areas
 across the United States.  Washington, D.C., the nation's capital, has been consistently
 ranked among the top cities that suffers from high rates of violent crimes. The impact of
 such violent crimes on the city are many fold, extending from residents to tourism, and
 the ability to curb such crimes is of utmost importance. In particular, there is
 significant interest in being able to predict regions in which crimes are likely to occur
 so that protective measures may be employed both in the short- and long-term.  Violent
 crimes often exhibit both temporal and spatial characteristics. The spatial patterns of
 crime do not vary smoothly across the map, and instead we see spatially disjoint areas
 that exhibit similar crime behaviors. It is this indeterminate inter-correlation
 structure of the multiple time series along with the low-count discrete nature of the
 data that motivate our proposed forecasting tool. In particular, we propose a novel
 methodology which models each time series as an integer-valued first order autoregressive
 process. We take a Bayesian nonparametric approach to discover how to spatially cluster
 the time series in a flexible, data-driven way. We illustrate our method's forecasting
 abilities both on simulated data and reported violent crime data in Washington, D.C.,
 collected between 2001-2008. These analyses show that our method outperforms standard
 methods and can also easily provide useful tools such as prediction intervals. We further
 demonstrate how one can modify our method to incorporate covariates such as population
 density and discuss the utility of this alternate formulation in the context of our data.
  }

 There is significant interest in being able to predict where crimes will happen, for example to aid
 in the efficient tasking of police and other protective measures. 
 We aim to model both the temporal and spatial dependencies often exhibited by violent crimes in order to make such predictions.
  The temporal variation of crimes typically
 follows patterns familiar in time series analysis, but the spatial patterns are irregular
 and do not vary smoothly across the area.  Instead we find that spatially disjoint
 regions exhibit correlated crime patterns.  It is this indeterminate inter-region
 correlation structure along with the low-count, discrete nature of counts of serious
 crimes that motivates our proposed forecasting tool. In particular, we propose to model
 the crime counts in each region using an integer-valued first order autoregressive
 process. We take a Bayesian nonparametric approach to flexibly discover a clustering of
 these region-specific time series.  We then describe how to account for covariates within this framework.  Both approaches adjust for seasonality. We demonstrate our approach through an analysis of weekly reported
 violent crimes in Washington, D.C. between 2001-2008.  Our forecasts outperform standard
 methods while additionally providing useful tools such as prediction intervals.  \\

{\small Keywords: Violet crime counts, Low-count time series, INAR, Bayesian nonparametric methods}
\end{abstract}
\doublespacing
\section{Introduction}\label{chapter:Introduction}
 Violent crimes are a significant concern in metropolitan areas across the United States.
  The impact of violent crimes on a city are many fold, ranging from harm to residents to
 loss of tourism, and the ability to curb such crimes is of utmost importance. In
 particular, there is significant interest in being able to predict regions in which
 crimes are likely to occur so that preventive measures may be employed in both the short-
 and long-term.  Predicting crimes relies on the presence of dependence over time and
 among regions.  The structure of the temporal dependence we find is familiar; for
 example, violent crimes are known to vary seasonally, with the rate of violent crimes
 higher during warmer months of the year \citep{mcdowall_seasonal_2011}. The spatial
 dependence is more complex. One might expect, for instance, neighboring regions to
 experience similar crime rates.  Geographic features, however, can impede crime from
 varying smoothly across regions.  For example in Washington, D.C., Rock Creek Park in the
 northeast and the Anacostia River in the south create natural impediments to the spread
 of crime. Boundaries also occur at a finer resolution (e.g. railroad tracks and
 highways).  The presence of such idiosyncratic features makes it challenging to devise
 statistical models that are able to borrow strength among similar local regions without
 over-smoothing distinct regions.
%

 As an example of our methodology, we consider rates of violent crimes in the 188 census
 tracts in Washington, D.C..  The nation's capital has consistently ranked among the top
 cities for rates of violent crimes in the United States. The crimes in our data consist of rape, robbery, arson and aggravated assault. Along with murder, these types of crimes define the FBI part 1 violent crimes list. Part 1 crimes are considered serious and are directly reported to the police (as opposed to other law agencies such as the IRS). Indeed, the Washington, D.C. police department keeps a record of all reported type 1 violent crimes and makes it publicly available through their website (\texttt{http://crimemap.dc.gov/CrimeMapSearch.aspx}). 
 
\begin{figure}[h!]
\centering
\begin{tabular}{cc}
\includegraphics[width = 1.9in]{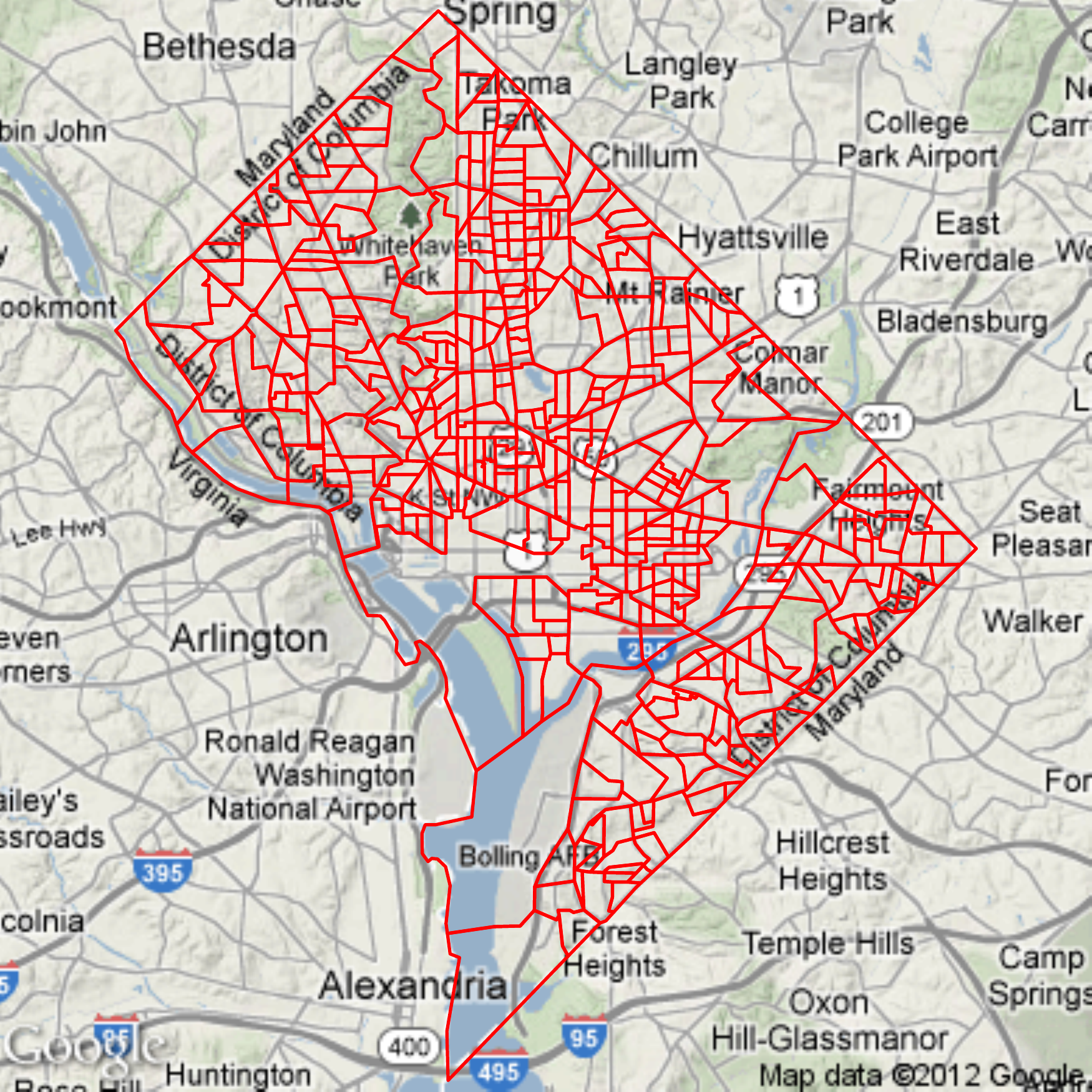} &\hspace{0.2in} \includegraphics[width=3.35in]{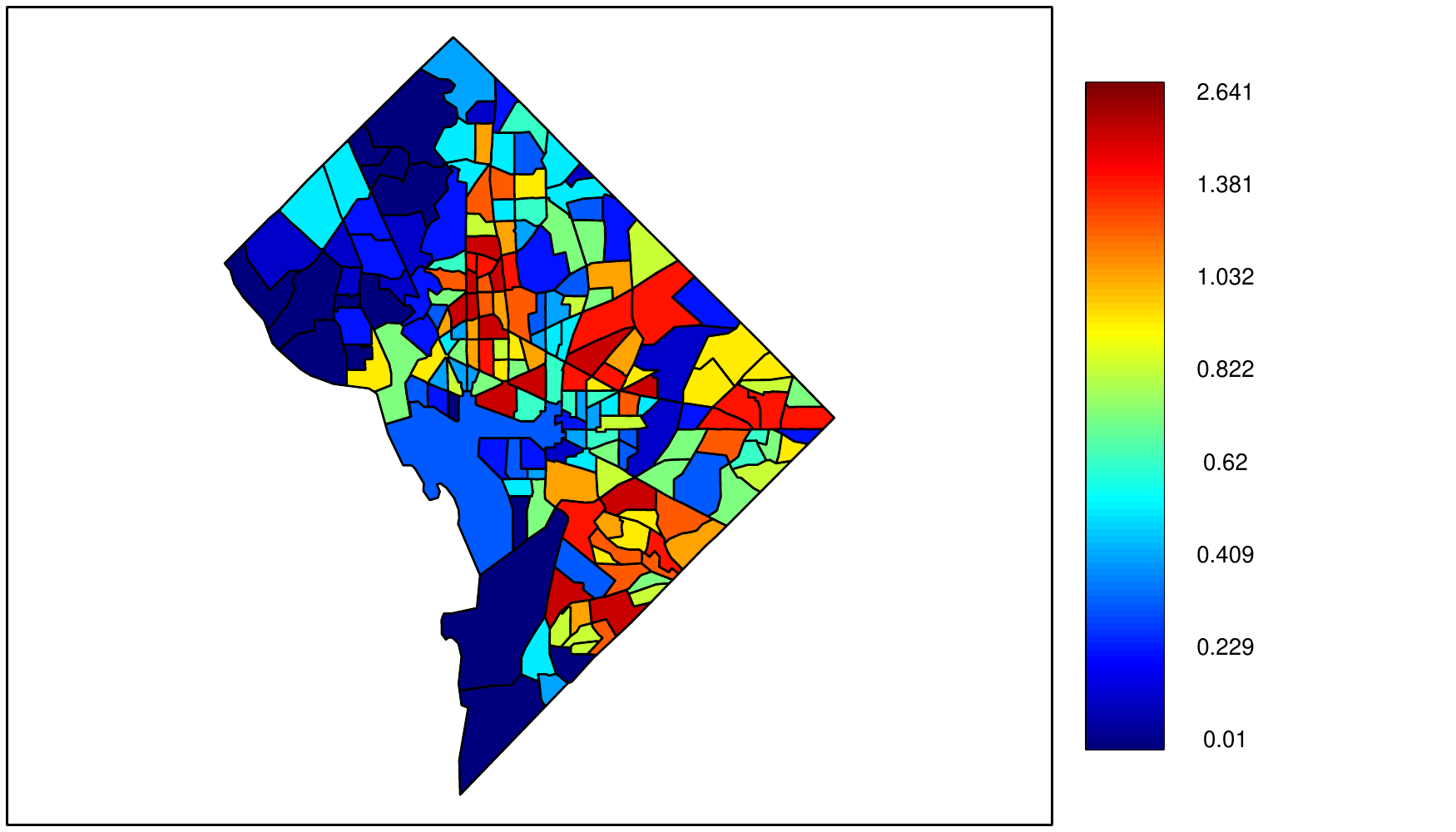}
\end{tabular}
\caption[Washington, D.C. 188 census tracts map.]{\emph{Left:} Map of the 188 census tracts in Washington, D.C.. \emph{Right:} Weekly average violent crime counts across the 188 census tracts.}
\label{fig:DC}
\end{figure}

 Figure \ref{fig:DC} (left) shows a map of Washington, D.C. with boundaries of the census tracts
 superimposed. The sizes of the census tracts vary widely depending on population density.
  A census tract consists of adjacent street blocks selected to be homogeneous with
 respect to demographic features such as economic status and living conditions.  According
 to the 2000 Census, tracts in Washington, D.C. average 3043 residents, ranging from 149 to 7278.
 Due to the demographic homogeneity of census tracks, it is not surprising that
 neighboring tracts often have different crime dynamics.

 The combination of demographic heterogeneity and natural boundaries contributes to the
 spatially diverse crime patterns across the census tracts. Figure \ref{fig:DC} (right)
 shows the weekly average counts of violent crimes in Washington, D.C. between 2001 and
 2008. This plot reveals several interesting features. First, the weekly average number of
 violent crimes are (fortunately) low. Second, tracts with similar average crime counts
 are often quite spatially disjoint. This suggests that crime, as hypothesized, does not
 vary smoothly across the city.

\begin{figure}[h!]
\centering
\includegraphics[scale=0.8]{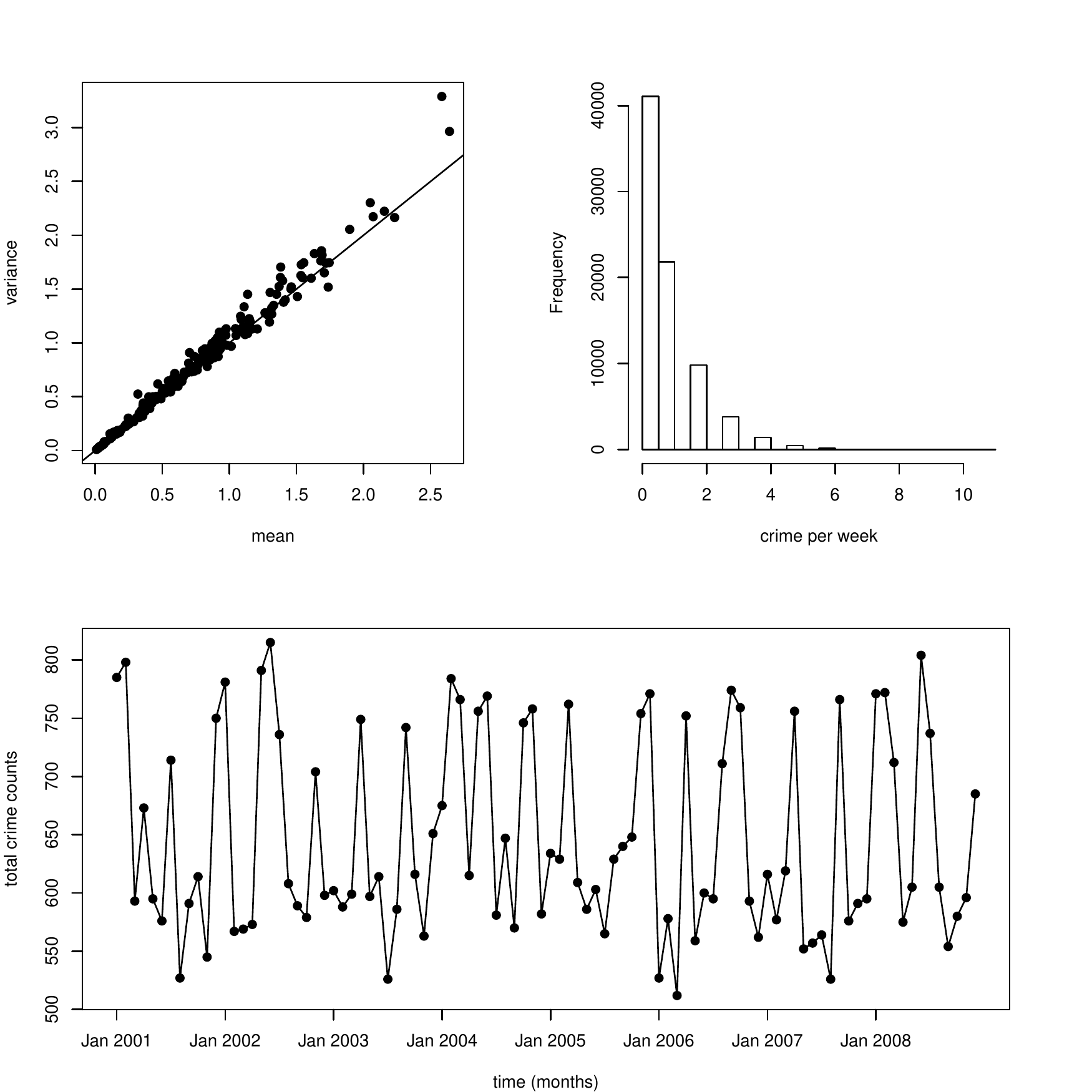}
\caption[Washington, D.C. 188 census tracts map.]{\emph{Top Left:} Histogram of weekly crime rates in Washington, D.C.'s 188 census tracts. \emph{Top Right:} Weekly standard
 deviation versus average violent crime counts across the 188 census tracts.  \emph{Bottom
 Right:} Total monthly crime counts.}
\label{fig:DCts}
\end{figure}

 The seasonal structure of weekly crime counts is difficult to see in sequence plots of
 the data for a single tract because of the small counts.  The pattern becomes evident,
 however, in the aggregate.  Figure \ref{fig:DCts} displays monthly crime counts
 accumulated over all of the tracts.  This plot reveals the seasonal peaks and valleys.
  (Supplementary Material Section~\ref{app:DCtime} includes plots of the counts for individual tracts.)
  Figure \ref{fig:DCts} also shows summary statistics accumulated over time for the
 individual tracts.  In particular, it shows the distribution of the total counts per
 tract over these years and a graph of the within-tract standard deviation versus the mean
 for all tracts.

The features of the weekly crime time series, namely (i) low-count discrete data with (ii)
 an uncertain correlation structure, necessitate the development of new forecasting
 tools. In particular, we propose a novel methodology which models each regional time
 series as an integer-valued first-order autoregressive process (INAR(1))
 \citep{alzaid_firstorder_1988,mckenzie_discrete_2000}.  We induce correlation
 between the time series through the innovations of the INAR(1) models.  We decompose
 these innovation processes into two latent factors: a shared seasonal effect and a rate
 function which is tract specific.  We use multiple shrinkage to cope with the large
 number of tracts.  We obtain this shrinkage adaptively by adopting a nonparametric
 Bayesian approach that imposes a Dirichlet process prior on the tract rates. The
 Dirichlet process leads to a clustering of rates and thus efficient sharing of the
 information between tracts in a flexible, data-driven manner.


 We begin in Section~\ref{sec:dependentPoINAR} with a purely geographic approach that is
 aligned with police procedures and work solely with raw crime counts within census
 tracts.  We then examine how to account for covariates, in particular population size, in
 Section~\ref{subsection: cov DoPoinar}.

We develop an efficient Markov chain Monte Carlo scheme for fitting these models, and we
 use this scheme to fit our model to simulated and real data.  The results demonstrate
 that our proposed multivariate INAR(1) model produces out-of-sample forecasts that are
 more accurate than those of a conditional least-squares (CLS) model in both simulations
 and when predicting crimes in Washington, D.C..  One reasonable explanation as to why our
 model outperforms CLS is because the number of clusters discovered is small relative to
 the number of tracts. Our model essentially shrinks the time series estimators that share
 a rate toward a common mean, thereby yielding better out-of-sample forecasts. Another
 advantage, a byproduct of the Bayesian framework, is that the Bayesian model provides
 posterior distributions of the $p$-step-ahead forecasts. These distributions are
 important in the context of forecasting crime because the distribution of crime is
 right-skewed and decision makers often care about preparing for the worst-case
 scenario. Prediction intervals for the number of violent crimes in each region can help
 the police distinguish between an unusual rise in violent crimes that requires
 intervention and a rise which is due to random variation.
\subsection{Related Approaches}
Multivariate Poisson-based models are a natural match to spatio-temporal time series of
 counts, and this structure has been employed in various prior applications.  For example,
 \citet{boudreault_multivariate_2011} model the occurrences of earthquakes using a
 maximum likelihood approach to infer the model parameters of a multivariate INAR(1)
 process with Poisson innovations.  (This formulation does not maintain Poisson margins,
 as discussed in Section~\ref{sec:background}.)  \citet{taddy_autoregressive_2010}
 employed Poisson processes to track the intensity of violent crimes in Cincinnati and treats these as point processes. 
 \citet{taddy_autoregressive_2010} factors the spatial Poisson rate into a process density, modeled using Bayesian
 nonparametrics, and an overall intensity.  Both were allowed to evolve in time.  Such a
 formulation, however, assumes spatial smoothness to the crime rates.  Additionally, 
 \citet{taddy_autoregressive_2010} focus on in-sample inference rather than on
 predicting future events.  In contrast, our research focuses on areal data modeling and provides methods to
 forecast these as multiple integer-valued low-count time series. We harness the efficient and
 elegant structure of INAR(1) processes and present a method for modeling multiple,
 correlated time series while maintaining Poisson margins.  The correlations are induced
 via a Bayesian nonparametric clustering of the time series, and in doing so, we
 efficiently share information to produce more accurate out-of-sample
 predictions. Bayesian nonparametric methods have previously been studied as tools for
 data-driven clustering analysis \citep[cf.,][]{teh_hierarchical_2006,
 dorazio_modeling_2008,fox_bayesian_2010}. However, these studies focus on clustering
 either continuous-valued time series or Poisson counts which have no time component.

Our paper is structured as follows. In Section~\ref{sec:background}, we provide background on the Poisson INAR(1).  We then present our method for correlating multiple such processes while maintaining Poisson margins in Section~\ref{sec:dependentPoINAR}.  Here, our focus is on correlation structure captured by geography alone.  The associated posterior computations via Markov chain Monte Carlo are detailed in Section~\ref{section:MCMC}.  In Section~\ref{sec:sim}, we briefly provide a simulation study and in Section~\ref{section:results}, we analyze our crime data of interest.  Finally, in Section~\ref{subsection: cov DoPoinar} we describe a method for accounting for covariates, and in particular consider population size as a predictor.  

\section{Univariate PoINAR(1) Background}
\label{sec:background}

 A univariate PoINAR(1) model is defined as follows \citep{alzaid_firstorder_1988}:
    \begin{equation}\label{Eq:inar}
        \obsVar_{t+1} =  \alpha \circ \obsVar_{t} + \epsilon_{t+1} \quad
           \textrm{for }t=0,1,2,\ldots \;,
    \end{equation}
 where the innovations $\{\epsilon_{t}\}$ are iid Poisson.  The operator $\circ$ denotes
 binomial thinning.  For any nonnegative integer-valued random variable $X$ and
 for any $\alpha \in [0,1]$, the random variable $\alpha \circ X$ is defined
    \begin{equation}\label{Eq:thin}
        \alpha \circ X = \sum_{i=1}^{X} B_i(\alpha),
    \end{equation}
 where $B_i(\alpha)$ are independent, identically distributed Bernoulli random variables
 with success probability $\alpha$.  Given a Poisson distribution on the initial state
 $\obsVar_0$ with finite mean $\mu = E\,\obsVar_0$ and independence between $\obsVar_t$
 and $\epsilon_t \sim \mbox{Poisson}((1-\alpha)\mu)$, the construction Eq.~\eqref{Eq:inar}
 yields a strongly stationary process.  In essence, to obtain a stationary Poisson
 marginal distribution from Eq.~\eqref{Eq:inar}, the innovations must also be Poisson
 \citep{alzaid_firstorder_1988, steutel_discrete_1986, larryes}.

%
%
 \hide{ We proceed by defining $\bm \obsVar_t :=(\obsVar_{1,t},\ldots,\obsVar_{L,t})$ as
 the multivariate violent crime counts during time $t$ at tracts $l=1,\ldots,L$ or
 alternatively one can view $\bm \obsVar_t$ as the map of crime counts at time $t$. To
 model $\bm \obsVar_{t}$ as a multivariate INAR(1) process,
 
 It is possible to extend Equations \ref{Eq:inar} and \ref{Eq:thin} to the multivariate case by 
 using a generalization of the binomial thinning operator.  Let $\bm X := (X_1,\ldots,X_L)$ denote a vector of
 nonnegative random variables and let $\bm \alpha$ denote an $L \times L$ matrix
 with entries in $[0,1]$. An $L$-dimensional thinning operator produces a
 $L$-dimensional random vector, with $i^\textrm{th}$ component
        \begin{align}
            [\bm \alpha \circ \bm X]_{i} : = \sum_{l=1}^L \alpha_{i,l} \circ X_{l},
        \end{align}
where $\alpha_{i,l} \circ X_l$ is the binomial thinning operator defined in
 \refeq{Eq:thin}.  The individual binomial thinning operators are assumed to be
 conditionally independent over indices, i.e. $\alpha_{i,l} \circ X_{l} \perp
 \alpha_{j,m} \circ X_{m} \| X$ for all $i,l,j,m$.
 
 Our application requires Poisson marginal distributions for all $t$, and this condition
 introduces restrictions on the multivariate INAR(1) process.  Let $\bm \epsilon_{t}:=
 (\epsilon_{1,t},\epsilon_{2,t},\ldots,\epsilon_{L,t})$ denote the vector of innovations
 at time $t$.  We can then extend the scalar model (1) to vectors as
 \begin{displaymath}
    {\bm \obsVar}_{t+1} = \bm \alpha \circ \bm \obsVar_{t} + \bm \epsilon_{t+1} \;.   
 \end{displaymath}
 Without further restrictions, this specification need not produce Poisson
 margins even if $\epsilon_{l,t}$ are independent Poisson random variables.
 In fact, it is straightforward to prove
 that when the off-diagonal elements of the thinning matrix $\bm \alpha$ are
 non-zero, a stationary distribution exists but is no longer the Poisson
 distribution \citep[see][]{mckenzie_discrete_2000,pedeli_bivariate_2011-1}.
  Such a multivariate INAR(1) was considered
 by~\citet{boudreault_multivariate_2011}.}

\section{Multivariate PoINAR(1)}
\label{sec:dependentPoINAR}
 In this section we define a multivariate INAR process that retains Poisson margins.  We
 first introduce the basic model and then demonstrate how to induce correlations among the
 component series by placing a Dirichlet process prior on the rate parameters of the
 Poisson innovations.  We conclude this section by highlighting the similarities and
 differences between the proposed model and the vector autoregressive process, which is
 the equivalent model with Gaussian margins.

Throughout, let $\obsVar_{l,t}$ denote the number of violent crimes at tract $l=1,\ldots,L$
 during week $t=1,\ldots,T$.  Furthermore, let $\bm \obsVar_t := (\obsVar_{1,t},\ldots,\obsVar_{L,t})$ denote a vector of crime counts at time $t$ and $\bm \epsilon_{t}:= (\epsilon_{1,t},\epsilon_{2,t},\ldots,\epsilon_{L,t})$ the vector of innovations.
\subsection{A multivariate PoINAR(1) process}
One might imagine employing a multivariate PoINAR(1) analogue of a vector autoregressive process by simply considering:
\begin{displaymath}
   {\bm \obsVar}_{t+1} = \bm \alpha \circ \bm \obsVar_{t} + \bm \epsilon_{t+1} \;.   
\end{displaymath}
where $\alpha$ is an $L \times L$ matrix with entries in $[0,1]$ and $\bm \alpha \circ \bm \obsVar_t$ is defined as:
\begin{align}
    [\bm \alpha \circ \bm \obsVar]_{i,t} : = \sum_{l=1}^L \alpha_{i,l} \circ Y_{l,t},
\end{align}
 with $\alpha_{i,l} \circ Y_{l,t}$ defined by the binomial thinning operator
 \refeq{Eq:thin}.  Even in the simplest scenario of $\epsilon_{i,t}$ being independent
 Poisson innovations, however, the resulting margins are in general not Poisson.  In fact,
 it is straightforward to prove that when the off-diagonal elements of the thinning matrix
 $\bm \alpha$ are non-zero, a stationary distribution exists but is no longer the Poisson
 distribution \citep[See][]{mckenzie_discrete_2000,pedeli_bivariate_2011-1}. 
 Such a multivariate INAR(1) was
 considered in~\citep{boudreault_multivariate_2011}.

 The one scenario in which Poissonicity is maintained is if $\bm \alpha$ is diagonal, {\it
 i.e.} $\alpha_{i,j}=0$ for $i \neq j$, so that the model becomes
\begin{align}\label{eq:multiInar}
\left( \begin{array}{c}
\obsVar_{1,t+1}\\
\obsVar_{2,t+1} \\
\vdots \\
\obsVar_{L,t+1} \end{array} \right) & =&
 \left( \begin{array}{cccc}
\alpha_1 & 0  &\ldots & 0 \\
0 & \alpha_2 & \ldots & 0 \\
\vdots & \vdots & \ddots & \vdots \\
0 & \ldots  & 0 & \alpha_L
\end{array} \right) \circ
\left( \begin{array}{c}
\obsVar_{1,t}\\
\obsVar_{2,t} \\
\vdots \\
\obsVar_{L,t} \end{array} \right) +
\left( \begin{array}{c}
\epsilon_{1,t+1}\\
\epsilon_{2,t+1} \\
\vdots \\
\epsilon_{L,t+1} \end{array} \right).
\end{align}

\noindent
For notational convenience we will denote the diagonal elements by
 $\alpha_{i}:=\alpha_{i,i}$.  The diagonal thinning matrix implies that at time
 $t$, the $l^\textrm{th}$ entry of the thinned random vector is only a function
 of only the $l^\textrm{th}$ tract:
\begin{align}
   [\bm \alpha \circ \bm \obsVar_t]_{l} = \alpha_{l} \circ \obsVar_{l,t}.
\end{align}

 For the innovations processes, we assume $\epsilon_{l,t}|\Lambda_{l,t} \sim
 \textrm{Poisson}(\Lambda_{l,t})$.  That is, conditional on the rate parameters,
 the innovations are independently Poisson distributed across time and
 space. The resulting multivariate INAR(1) yields a stationary Poisson
 distribution for each element in $\mathbf{\obsVar}_t$.  We refer to this
 process as the \emph{multivariate PoINAR(1)}.  We emphasize that the diagonal
 thinning matrix not only dramatically reduces the number of model parameters,
 but also produces a stationary process with Poisson  margins.

 Conditioning on the rate parameters $\{\Lambda_{l,t}\}$ yields $L$ independent time
 series.  To allow such models to capture dependencies between the time series, we
 introduce a Dirichlet process mixture model for the innovations in the next section.
\subsection{Capturing dependencies}\label{subsection:DPoinar}
There are several ways to induce dependencies between the elements of the
 multivariate PoINAR(1) process. As previously mentioned, there are two sources
 of variation in the model: the multivariate binomial thinning operator and the
 innovation process. We propose to generate the dependence through the
 innovations and assume that the binomial thinning operators are independent
 across the time series.  For our model of crime rates, this formulation shares
 information between tracts while allowing tract-dependent
 autocorrelations. Furthermore, this focus on the innovations provides
 computational efficiencies as described in Section \ref{section:MCMC}.

Recall that the innovations are assumed to follow a Poisson distribution with
 rates $\Lambda_{l,t}$. The rate is a function of both the tract $l$ and the
 time period $t$ of the specific innovation $\epsilon_{l,t}$. We decompose the
 rate $\Lambda_{l,t}$ into a product of a spatial and temporal components:
 \begin{align}
      \Lambda_{l,t} = \lambda_l \, \theta_{s(t)},    
 \end{align}
 where the summands are
\begin{itemize}
  \item a tract-specific rate, $\tractRate_l$, and
  \item a seasonal monthly rate, $\theta_{s(t)}$, that is spatially homogeneous.  Here,
 $s(t)$ is a function that maps week $t$ to its associated month.  That is, we assume a
 constant seasonal effect within months and model this effect with 12 parameters
 ${\theta_1,\ldots,\theta_{12}}$.  
\end{itemize}
 The resulting model for the innovations can be written as follows:
    \begin{eqnarray}
        \epsilon_{l,t} &\sim& \textrm{Poisson}(\tractRate_l \, \theta_{s(t)}).
    \end{eqnarray}
 The temporal component induces some dependence across tracts because it is
 shared across the different time series.  A Dirichlet process (DP) prior on the
 rates, $\tractRate_l$, provides the balance of the dependence.

 The Dirichlet process, denoted DP$(\tau,G_0)$, provides a distribution over countably
 infinite probability measures.  $G_0$ denotes a probability distribution over some space
 $\Omega$; in our application, $\Omega$ is the positive real line.  The concentration
 parameter $\tau > 0$ controls the `clumpyness' of the process.  
 A Dirichlet process can be written as a weighted sum of point masses
 $\delta_x$ positioned by randomly sampling $G_0$:
  \begin{align}\label{eq:DP}
     \textrm{G} &= \sum_{k=1}^{\infty} \beta_k \, \delta_{\uniqueRate_k} ,
           \qquad \uniqueRate_k \stackrel{iid}{\sim} G_0.
  \end{align}
 The weights $\beta_k$ can be obtained sequentially via the so-called
 \emph{stick-breaking} construction \citep{sethuraman_constructive_1994}:
  \begin{align}\label{eq:SBP}
    \beta_k &= \nu_k \prod_{l=1}^{k-1} (1-\nu_l) 
         \qquad \nu_k \sim \textrm{Beta}(1,\tau).
  \end{align}
 In effect, the process divides the unit interval into segments with lengths
 given by the decreasing sequence of weights $\beta_k$: the $k^{\textrm{th}}$
 weight is a random proportion $\nu_k$ of the segment that remains after the
 first $k-1$ weights have been chosen.  We denote this distribution by
 $\beta\sim \mbox{GEM}(\tau)$.  

 The DP has proven useful in many applications due to its clustering properties 
 \citep[cf.,][]{teh_hierarchical_2006}. The \emph{predictive distribution} of draws $W \sim \textrm{G}$ shows why
 the DP produces clusters.  Because probability measures drawn from a DP are discrete by
 construction, there is a strictly positive probability of multiple observations of $W$
 taking identical values within the set $\{\uniqueRate_k\}$, with $\uniqueRate_k$ defined
 as in \refeq{eq:DP}.  For each sampled observation $w_i$, let $z_i$ identify the position
 of the corresponding parameter $\uniqueRate_k$ such that $w_i = \uniqueRate_{z_i}$.  The
 predictive distribution on the membership variables can be written in the following
 manner:
%
   \begin{align}\label{eq:CRP} 
	Z_{N+1} |(z_1,\ldots,z_N,\tau)  = \left \{ \begin{array}{ll}
         K+1 & \mbox{w.p. $  \frac{\tau}{N+\tau}$}\\
        k & \mbox{w.p. $ \frac{n_k}{N+\tau}$ for $k=1,\ldots,K$},\end{array} \right. 
\end{align}   
%
where $n_k$ indicates the number
 of members belonging to the $k^\textrm{th}$ group and $K = \max\{z_1,\ldots,z_N\}$
 identifies the number of distinct values observed through the first $N$ samples.  The
 distribution on partitions induced by the sequence of conditional distributions in
 Eq.~\eqref{eq:CRP} is commonly referred to as the \emph{Chinese restaurant process}
 (CRP). The CRP provides an alternative representation to the DP \citep{pitman}. This
 representation emphasizes the reinforcement property of the DP that leads to its
 clustering properties.  It can be shown that the expected number of clusters using the
 CRP grows as $O(\tau \, \textrm{log}(L))$ where $L$ is the number of observations \citep[see][for a detailed proof.]{sammut_dirichlet_2011}. This implies that the average number
 of clusters is much smaller than the number of observations.
 
 In our model for crime rates, we impose a DP prior on the $L$ tract-specific rates,
 $\tractRate_l$.  Note that in our application, the number of observations from the DP is
 equal to the number of tracts, rather than the number of time points.  The DP prior thus
 groups the time series according to their corresponding tract-specific rates into a few
 clusters.  Members of the $k^\textrm{th}$ cluster share a common tract rate,
 $\uniqueRate_k$.  The grouping of the time series into a small number of clusters
 provides useful shrinkage that pools information across the cluster, thereby yielding
 more accurate out-of-sample predictions for the multiple time series.  When we combine
 the tract-specific rates and the seasonal effects, we obtain the following generating
 process for the innovations:
\begin{align}
	\begin{aligned}
\epsilon_{l,t} &\sim \textrm{Poisson}(\tractRate_{l} \, \theta_{s(t)}) \quad l=1,\ldots,L \quad t=1,\ldots,T\\ 
 \theta_m & \sim  \textrm{F} \quad m=1,\dots,12\\ 
\tractRate_{l} &\sim \textrm{G} \quad l=1,\dots,L\\ 
 \textrm{G} &\sim  \textrm{DP}(\tau,G_0).
\end{aligned}
\label{eqn:DP1}
\end{align}
For our application, we choose $F$ to be a gamma distribution in Eq.~\eqref{eq:gammabase}.

\begin{figure}
\begin{tabular}{cc}\hspace{-0.5cm}
	\vspace{-0.2in}
\includegraphics[scale=0.1,width=2.8in]{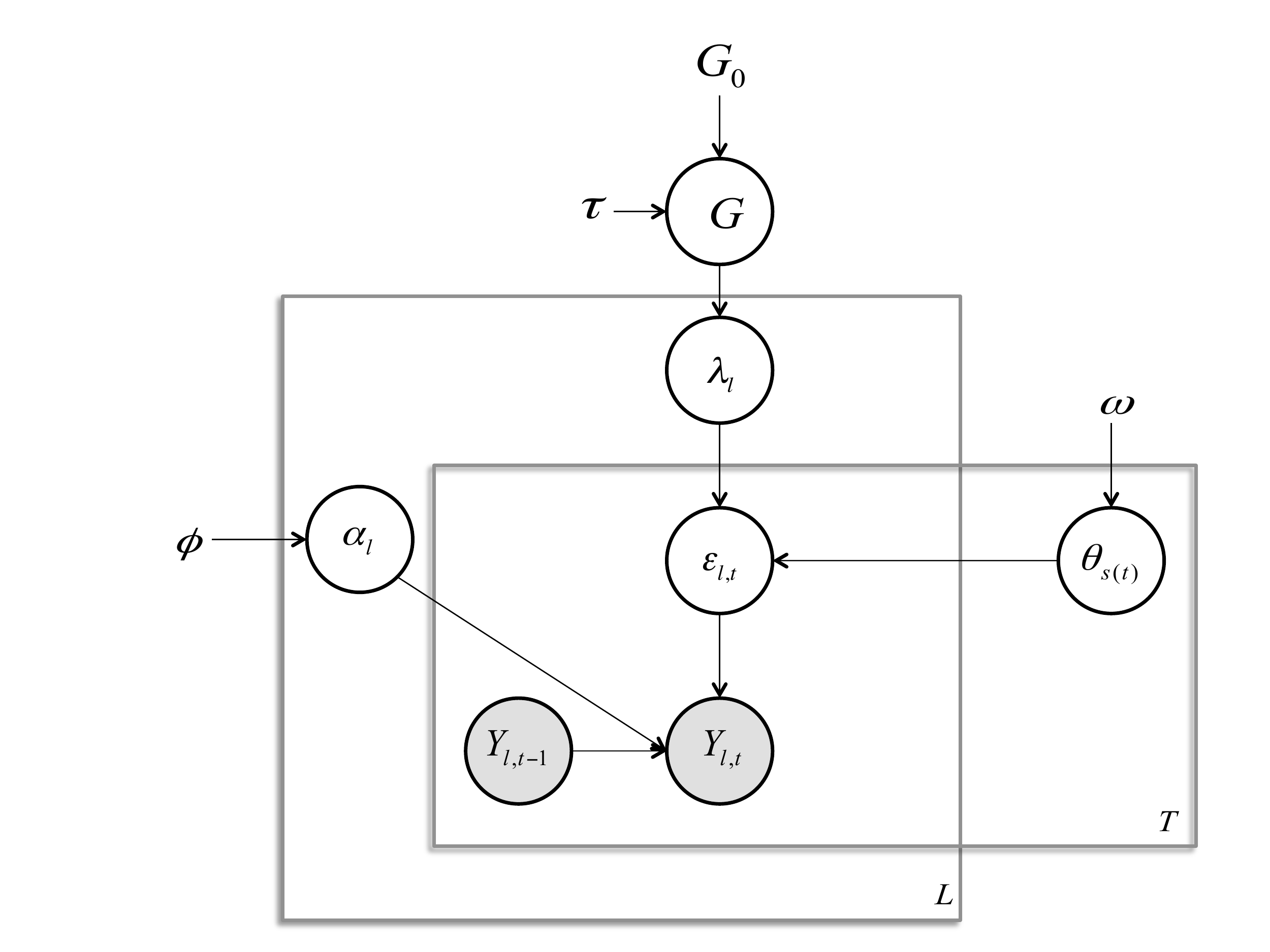} &
\includegraphics[scale=0.1,width=2.8in]{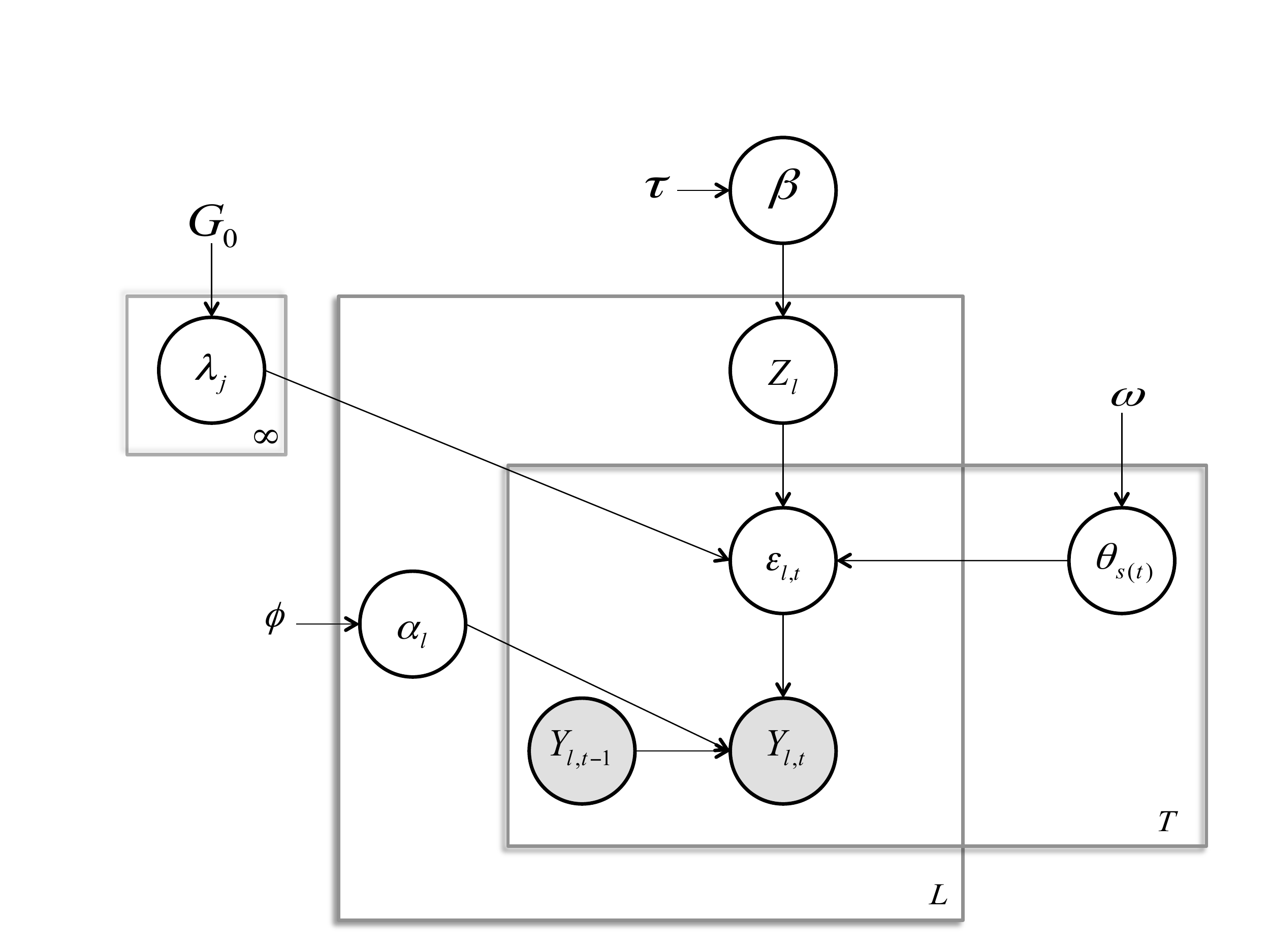} \\
\end{tabular}
\caption[The multivariate dependent PoINAR(1) model.]{Graphical model of the multivariate
 dependent PoINAR(1) model. \emph{Left:} Innovations
 generating from a Dirichlet process mixture model as in Eq.~\eqref{eqn:DP1}. \emph{Right:} An equivalent representation
 using cluster indicator variables $z_1,\ldots,z_L$ as in Eq.~\eqref{eqn:DP2}.}
\label{fig:DP}
\end{figure}
Figure \ref{fig:DP} (left) shows a graphical representation of our dependent multivariate
 PoINAR(1) process. For details on interpreting such graphical representations, the reader is referred to \cite{graphs}. 
 Alternatively, we can use an equivalent representation using the GEM
 distribution and the membership labels $z_1,\ldots,z_L$ (see Figure~\ref{fig:DP}
 (right)):
\begin{align}
	\begin{aligned}
\epsilon_{l,t} &\sim \textrm{Poisson}(\uniqueRate_{z_l} \, \theta_{s(t)}) \quad l=1,\ldots,L \quad t=1,\ldots,T\\ 
\theta_m & \sim \textrm{F} \quad m=1,\dots,12\\ 
z_l & \sim \textrm{Multinomial}(\beta) \quad l=1,\dots,L\\ 
\uniqueRate_j &\sim G_0 \quad \quad \beta \sim \textrm{GEM}(\tau) \quad j=1,2,\dots.
\end{aligned}
\label{eqn:DP2}
\end{align}
%
%

\subsection{Prior specification}
The multivariate PoINAR(1) requires estimation of three main components:
    \begin{itemize}
        \item thinning values $(\alpha_1,\ldots,\alpha_L)$, one for each time series.
        \item monthly seasonal effects, $(\theta_1,\ldots,\theta_{12})$.
        \item rates for each tract, $[\tractRate_1,\ldots,\tractRate_L]$.
    \end{itemize}

 The Bayesian framework places prior distributions on each of these three elements.  Our
 priors are both computationally convenient and weakly-informative. 
For the thinning values and monthly
 seasonal effects we specify:
            \begin{align}
                \alpha_l &\iid \textrm{Beta}(\eta_1,\eta_2) \quad \textrm{for } l=1,\ldots,L \\
             \nonumber   \theta_m &\iid \textrm{Gamma}(\xi_1,\xi_2) \quad \textrm{for } m=1,\ldots,12.
              \end{align}
 We also explored the half-normal distribution as a prior for the seasonal effect, and we
 found the choice did not reveal any material changes from the results presented in
 Section~\ref{section:results}.  The DP requires the specification of the base measure,
 $G_0$, and the concentration parameter, $\tau$.  We choose the base measure to be the
 Gamma$(\gamma_1,\gamma_2)$ distribution, which is well suited to our model because not
 only it is conjugate to the Poisson distribution but it also provides a natural
 interpretation.  In particular, we have a prior belief that weekly rates of violent crime
 are typically small, but a few tracts have higher rates.  Hence, a gamma
 distribution with shape and scale parameters $\gamma_1=1$ and $\gamma_2=0.1$ reflects
 these prior beliefs.  For the concentration parameter, we specify $\tau \sim
 \mbox{Gamma}(a_{\tau},b_{\tau})$, as suggested by \citet{escobar_bayesian_1994}.

\subsection{Relationship to Multivariate Independent AR(1) process}
The continuous counterpart to the multivariate PoINAR(1) is the Gaussian multivariate first-order autoregressive process (VAR(1)).  
This process is composed from $L$ independent AR(1) processes and can be formulated in the following manner:
\begin{align}
	\begin{aligned}
\bm \obsVar_{t+1} &= \bm \alpha \cdot \bm \obsVar_t +\bm \epsilon_{t+1} \quad t=1,\ldots,T\\
\bm \epsilon_{t} | \Sigma & \iid  N(0,\Sigma)\\
\obsVar_{l,0}|\mu_{l,0},\sigma_{l,0} & \iid  N(\mu_{l,0},\sigma_{l,0}) \quad l=1,\ldots,L.
\end{aligned}
\end{align}
where $[\Sigma]_{i,j}=0$ for $i \neq j$ and $[\Sigma]_{i,i}=\sigma^2_{i}$. 
Compare this specification to the multivariate PoINAR(1):
\begin{align}
	\begin{aligned}
\bm \obsVar_{t+1} &=  \bm \alpha \circ \bm \obsVar_t +\bm \epsilon_{t+1} \quad t=1,\ldots,T\\
\bm \epsilon_{t}|\Lambda_{t} & \iid  [\textrm{Poisson}(\Lambda_{1,t}),\textrm{Poisson}(\Lambda_{2,t}),\ldots,\textrm{Poisson}(\Lambda_{L,t})] \\
\obsVar_{l,0}|\Lambda_{l,0} &\iid   \textrm{Poisson}(\Lambda_{l,0}) \quad l=1,\ldots,L.
\end{aligned}
\end{align}
 These two models not only share similar notation, but also possess two common
 characteristics:
\begin{itemize}
  \item The distributions of the innovations match the marginal distributions
 of $\obsVar_{l,t}$.  The PoINAR(1) with diagonal $\bm \alpha$ (having non-negative diagonal entries) 
 has Poisson marginal distributions while the VAR(1) has Gaussian marginals for any $\bm \alpha$.
 If the parameters defining the processes are chosen carefully, these models can
 be shown to have a Poisson or Gaussian stationary distribution, respectively.
  \item The autoregressive parameter $\alpha_{l,l}$ determines the autocorrelation
 coefficient in both models, $\textrm{corr}(\obsVar_{l,t+1},\obsVar_{l,t})=\alpha_{l,l}$.
\end{itemize}

These similarities to the continuous VAR(1) make the discrete PoINAR(1) especially
 attractive and easy to interpret.  The VAR(1) process, however, has a single source of
 variation -- the innovations process -- while the PoINAR(1) process has two -- the
 binomial thinning and innovations processes.  This key difference complicates inference
 for the PoINAR(1) model, which we address in Section~\ref{section:MCMC}.
\section{The MCMC Sampler}\label{section:MCMC}
As previously noted, the PoINAR(1) model is a combination of two underlying processes: the
 binomial thinning process and the innovations process.  Each of these processes has its
 own parameters: binomial thinning uses the thinning parameters $\{ \alpha_l \}$ and the
 innovations process uses the rates $\{\uniqueRate_k\}$ and the seasonal effects $\{
 \theta_m\}$.  For posterior computations within our Bayesian framework, we employ an MCMC
 sampler. Intuitively, the idea is to sample a posterior latent innovations sequence and
 then condition on this sequence to sample both the latent DP clustering of census tracts
 and also the thinning parameters and seasonal effects.  In contrast, in the corresponding
 VAR(1) model there is no need to sample the innovations sequence since they are
 a deterministic function of the observations and the model parameters.  Therefore, one
 would expect the multivariate PoINAR(1) model to be computationally cumbersome compared
 to its VAR(1) counterpart.  However, our proposed sampler harnesses computational
 advantages from small observed counts in our crime data and sufficient statistics implied
 by the Poisson model. We outline the resulting sampler below.  For detailed derivations,
 see Section~\ref{section:appendix mcmc} of the Supplementary Material.
              \begin{enumerate}
                \item \label{step:error sample} Sample the innovations, $\bm \epsilon:= [\epsilon_{1},\ldots, \epsilon_{L}]$ where $\epsilon_l:= [\epsilon_{l,1},\ldots,\epsilon_{l,T}]$ is the innovations series for the $l^{\textrm{th}}$ tract.  The posterior factors as:
                \begin{align}\label{eq:sample epsilon}
                P(\bm \epsilon|\bm \obsVar, \bm \alpha, \bm \tractRate,\bm \theta ) &= \prod_{l=1}^L \prod_{t=2}^T P(\epsilon_{l,t} | \obsVar_{l,t-1},\obsVar_{l,t},\alpha_l, \tractRate_l,\bm \theta).
                \end{align}
align
Given the observations $\bm \obsVar$ and the parameters of the multivariate PoINAR
 process, the innovations can be sampled independently for each tract and each time
 point. The possible values satisfy $\max\{0,\obsVar_{l,t}-\obsVar_{l,t-1}\}\leq
 \epsilon_{l,t} \leq \obsVar_{l,t}$
       with corresponding probabilities
  \begin{multline}
  P(\epsilon_{l,t} | \obsVar_{l,t-1},\obsVar_{l,t},\alpha_l, \tractRate_l,\bm \theta)
\propto\\
 \frac{1}{\epsilon_{l,t}!(\obsVar_{l,t}-\epsilon_{l,t})!(\obsVar_{l,t-1}-(\obsVar_{l,t}-\epsilon_{l,t}))!}
  \left (\frac{\tractRate_l \theta_{s(t)} \, (1-\alpha_l)}{\alpha_l}
 \right)^{\epsilon_{l,t}} .
  \end{multline}
 \noindent
 Although this expression does not define a well-known discrete distribution, it is
 analytically tractable because of the small counts in the data (i.e.,
 $\max\{0,\obsVar_{l,t} - \obsVar_{l,t-1}\} \leq \epsilon_{l,t} \leq \obsVar_{l,t}$ and
 $\obsVar_{l,t}$ is assumed to be small).  In the crime data for Washington, D.C.,
 $\max{\obsVar_{l,t}}= 11$.  Another important consideration that reduces the
 computational burden is that certain $\epsilon_{l,t}$ values can be deterministically set
 from the observations vector $\bm \obsVar_t$: if $\obsVal_{l,t} = 0$ then
 $\epsilon_{l,t}=0$ and if $\obsVal_{l,t-1} = 0$ then $\epsilon_{l,t} = \obsVal_{l,t}$.
  Since our crime data has many zero counts, these constraints substantially lower the
 computational cost of this portion of the sampling.  If larger counts are observed, then
 one can use a Metropolis-Hastings step to sample from this distribution with a Poisson
 proposal distribution.  Importantly, note that if the observed counts are large enough
 one can apply a stabilizing transformation such as the square root and model the
 resulting process as a VAR. This strategy has been shown by \cite{CC} to yield
 satisfactory results in the univariate case when $\lambda_l$ is assumed to exceed
 (roughly) 5.

               \item \label{step:membership sample} Sample the membership indicator
 vector, $\bm z:=[z_1,\ldots,z_L]$. We harness the DP-induced Chinese restaurant process
 (CRP) and iteratively sample tract-specific cluster indicators:
                 \begin{align}
	P( z_{l}=k |\bm z_{/l},\bm \epsilon, \Theta, \gamma_1,\gamma_2,\tau) \propto
 \left\{ \begin{array}{ll} \tau \, p_{l,0} & \textrm{for } k = K+1 \\ n_k \, p_{l,k} &
 \textrm{for } k =1,\ldots,K,
                \end{array} \right.
				\label{eq:indpost}
			\end{align}
                where $K+1$ identifies a previously unseen cluster, $\Theta=\sum_{t=1}^T
 \theta_{s(t)}$ and $\bm z_{/l}$ is the vector of membership indicators, not including the
 $l^\textrm{th}$ term.  The first terms, $(\tau, n_j)$, of Eq.~\eqref{eq:indpost} arise
 from the CRP prior of Eq.~\eqref{eq:CRP} and the exchangeability of the process such that
 each $z_l$ can be treated as the last.  The second terms, $(p_{l,0},p_{l,j})$, correspond
 to the likelihood of the innovations $\bm \epsilon$ given the cluster assignments
 $(z_l=k,\bm z_{/l})$ and seasonal effects $\Theta$, marginalizing the cluster-specific
 rates $\uniqueRate_k$.  The terms are given by the following negative binomial
 distributions:
                \begin{align}
                p_{l,0} &= \frac{\Gamma(S_l +\gamma_1)}{\Gamma(\gamma_1) S_l!} \left ( \frac{\gamma_2}{\Theta+\gamma_2} \right)^{\gamma_1} \left ( \frac{\Theta}{ \Theta+\gamma_2} \right)^{S_l} \\
                \nonumber p_{l,j} &= \frac{\Gamma(S_l+A_j
+\gamma_1)}{\Gamma(A_j+\gamma_1) S_l!} \left ( 1- \frac{\Theta }{n_j \, \Theta+\gamma_2} \right)^{A_j +\gamma_1} \left ( \frac{\Theta }{n_j \, \Theta+\gamma_2} \right)^{S_l}, 
                \end{align}
                where $S_l = \sum_{t=1}^T \epsilon_{l,t}$ and $A_j=\sum_{i: z_i=j, i\neq
 l} S_i$. Note that $p_{l,0}$ and $p_{l,j}$ only rely on sums of the innovations
 and the sum of seasonal effects.  We also highlight that the conditional conjugacy of our
 formulation allows us to use the collapsed sampler of Eq.~\eqref{eq:indpost} for the
 $z_l$, marginalizing $\{\uniqueRate_k\}$.
                \item \label{step:rate sample} Sample unique rates,
 $\uniqueRate_k$. Although the rates  collapse away in sampling the cluster
 indicators, $z_l$, they are needed for sampling the innovations sequence (Step 1) and
 seasonal effects (Step 3).  As such, we instantiate the unique rates as auxiliary
 variables for these steps, and then discard them.  For each currently instantiated
 cluster, sample $\uniqueRate_k$ as:
                \begin{align}
                \uniqueRate_k |\bm \epsilon, \bm z, \Theta, \gamma_1, \gamma_2 &\sim \textrm{Gamma}(B_k+\gamma_1,n_k \, \Theta +\gamma_2),
                \end{align}
                where $B_k = \sum_{l\in \{v: z_v=k\}} S_l$. Again, we only rely on the sum
 of the innovations, $S_l$, to compute the posterior distribution.
                \item \label{step:seasonal sample} Sample the seasonal effects vector, $[\theta_1,\ldots,\theta_{12}]$.  The $m^\textrm{th}$ element of this vector can be sampled as:
                \begin{align}
                \theta_m | \bm \epsilon, \bm \uniqueRate, \xi_1, \xi_2 &\sim
\textrm{Gamma}\left(\sum_{l=1}^L \sum_{t:s(t)=m} \epsilon_{l,t} + \xi_1, q_m \, \sum_{l=1}^L \tractRate_l + \xi_2\right),
\label{eq:gammabase}
                \end{align}
                where $q_m$ counts the number of occurrences of the $m^\textrm{th}$
 month in the data. Notice that for this step we 
 sum the innovations over tracts rather than time.
                \item \label{step:thinning sample} Sample the vector of thinning
 parameters, $[\alpha_1,\ldots,\alpha_L]$.  For tract $l$,
                \begin{align}
                    \alpha_l | \bm \epsilon_l,\bm \obsVar_l, \eta_1, \eta_2 &\sim  \textrm{Beta}  (\sum_{t=2}^T \obsVar_{l,t}- S_l + \eta_1,\sum_{t=2}^T (\obsVar_{l,t-1} - \obsVar_{l,t}) + S_l + \eta_2),
                \end{align}
                where $S_l$ is defined as in Step 2.
                \item \label{step:concentration sample} Sample the concentration parameter, $\tau$, for the Dirichlet process prior according to \citet{escobar_bayesian_1994}.
                 \end{enumerate}

 It is important to note that if the model did not include seasonal effects,
 then one could simply sample the sum of the innovations, $S_l$, instead of the vector of
 innovations, $\epsilon_l$. This would reduce the computational cost of the sampler since
 Step 1 is the most time consuming.
\section{Simulation Examples} \label{sec:sim}
 In order to demonstrate the performance of our model, we simulate datasets from 9
 different multivariate PoINAR(1) processes of Eq.~\eqref{eq:multiInar}.  Each dataset has
 $L=100$ time series (tracts) with $T=208$ observations which correspond to 4 years of
 weekly data samples. We group the multiple time series into four equally sized clusters
 that each share a common rate.  The data sets vary in the choice of:
\begin{itemize}
\item The separation between the cluster rates, $\uniqueRate_k$.  We examine an
 ``easy'' setting in which the four cluster rates are well separated at $1, 3, 6, 10$, a
 ``medium setting'' with less distinct rates $0.01, 0.5, 1.2, 2$, and a ``hard'' setting
 with rates $0.1, 0.2, 0.3, 0.6$.
\item The thinning values, $\alpha_l$, which determine the autocorrelation of the
 individual PoINAR(1) processes. The examples use a common choice for $\alpha_l$ for all
 tracts in a dataset, choosing from $\alpha_l = 0.1, 0.5, 0.9$.
\end{itemize}

 We evaluate the root mean square error (RMSE) and absolute percentage error (APE) of our
 MCMC sampler both in- and out-of-sample.  These metrics measure the distance between the true
 population expected value and its corresponding estimate based on the observed $L=100$
 time series.  The simulation results show that our model produces accurate out-of-sample forecasts
 under various configurations.  Table \ref{table:SimRes2} presents the RMSE of our
 Bayesian nonparametric model compared to the RMSE of a simple Poisson process model (SPP)
 and the conditional least-squares model (CLS).  (The Supplementary Material, Section~\ref{Section:appendix cls},
 details both of these.)  The results in Table \ref{table:SimRes2} show that our model
 outperforms these alternatives.  As expected, the larger the separation between the
 cluster rates, the easier it is for our method to identify the true clusters and yield better estimates for their parameters.  
 Also, higher autocorrelation helps our method produce more accurate estimators.
\begin{table}[ht!]\footnotesize\centering
\begin{tabular}
{|l|c|c|c|c|c|c|c|c|c|}
\hline
& \multicolumn{3}{c}{Thin=0.1} & \multicolumn{3}{|c|}{Thin=0.5} &  \multicolumn{3}{c|}{Thin=0.9} \\
\hline
\hline
Rates & Easy & Med & Hard & Easy & Med & Hard & Easy & Med & Hard \\
\hline
SPP RMSE &  0.477 & 0.113 & 0.005 & 1.674 & 0.880 & 0.293 & 6.128 & 1.155 & 0.552\\
\hline
CLS RMSE & 0.306 & 0.080 & 0.035 & 0.284 & 0.114 & 0.057 & 0.343 & 0.118 & 0.055 \\
\hline
BNP RMSE & \textbf{0.219} &\textbf{0.058} & \textbf{0.026} & \textbf{0.260} & \textbf{0.086} & \textbf{0.045} & \textbf{0.299} & \textbf{0.075} & \textbf{0.043}  \\
\hline
$E(\obsVar_{\cdot,T+1})$ & 5.383 & 1.001 & 0.317 & 9.861 & 1.848 & 0.591 & 52.161 & 9.908 & 3.0633 \\
\hline
\end{tabular}
\caption[Estimation performance comparison between the SPP, CLS and MCMC methods]{RMSE of
 estimates of the conditional mean obtained by the CLS, SPP and our Bayesian nonparametric method.  The last
 row shows the expected (true) conditional expected value.}\label{table:SimRes2}
\end{table}

 These simulation results indicate that the sampler finds clusters when they exist.  It is also
 important to demonstrate that the model does not spuriously spawn clusters when the data are
 homogeneous.  As part of our simulations, we also examined the performance of these
 methods in the situation in which a single process (cluster) generates all of the time series.  The findings are presented in Section~\ref{app:sim} of the Supplementary Material, which also contains a more detailed description of the simulations and results presented in this section. 
As one would hope, under these conditions we identify a single cluster, further validating our methodology.
\section{Violent Crime Data Analysis}\label{section:results}
In this section we examine both in- and out-of-sample results using the reported counts of
 violent crimes in Washington, D.C. as described in Section \ref{chapter:Introduction}.  The data
 consist of $L=188$ time series (census tracts) with $T=418$ weeks of counts in
 2001 through 2008.  We use the first 7 years of data to train our model and the last 52
 weeks to evaluate its out-of-sample forecasts.  We ran 5 MCMC chains for $5,000$ iterations
 from different initial values, each drawn from the following priors: 
\begin{align}
	\begin{aligned}
  \theta_m &\sim \textrm{Gamma}(1,1) \quad \alpha_l \sim \textrm{Beta}(1,1) \\
  \tau &\sim \textrm{Gamma}(2,4) \quad \uniqueRate_i \sim \textrm{Gamma}(1,1).
	\end{aligned}
\end{align}
 We performed a sensitivity analysis for the hyperparameters during the simulation stage,
 but found no significant changes to the results. We discard the first 1000 iterations as
 burn-in and then thin the remaining $4,000$ samples by $50$. Therefore, our inference for
 each parameter of the model is based on the resulting $80 \cdot 5 = 400$ MCMC samples. We
 use the scale reduction factor \citep[recommended by][]{gelman_inference_1992} to monitor
 convergence across the chains.

\begin{figure}
\centering
\includegraphics[scale=0.8]{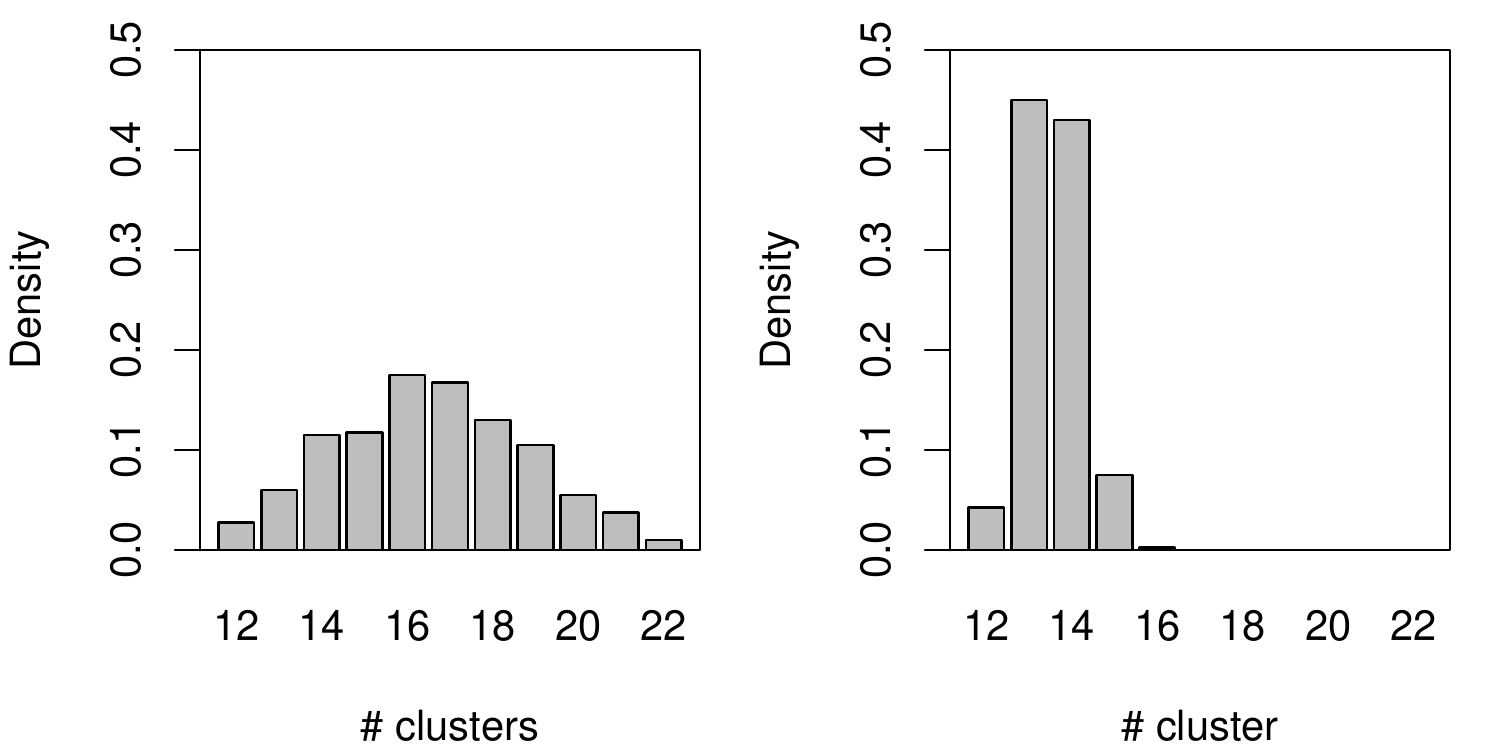}
\caption[Histogram of the posterior number of clusters.]{Histograms of the posterior number of clusters for the multivariate dependent PoINAR(1) described in Section \ref{subsection:DPoinar} (left) and the population adjusted multivariate dependent PoINAR(1) model described in Section \ref{subsection: cov DoPoinar} (right).}
\label{fig:ClusterHist}
\end{figure}

\begin{figure}
\centering
\includegraphics[scale=0.65]{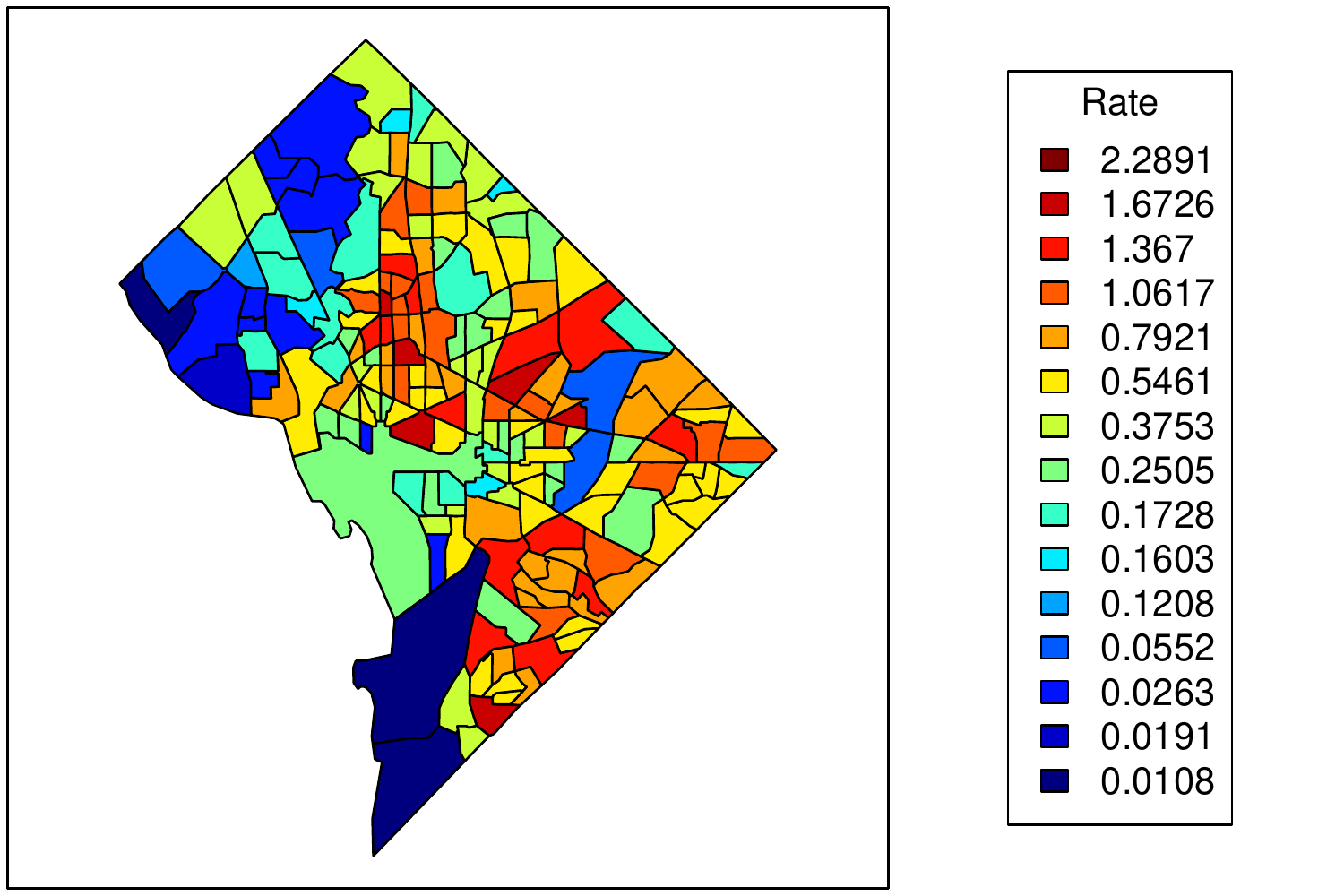}
\caption[The minimum average Hamming distance cluster assignment for the 188 Washington, D.C. census tracts.]{The minimum average Hamming distance cluster
assignment along with the corresponding posterior rate values.}
\label{fig:ClusterMap}
\end{figure}

We begin by looking at the distribution of the number of clusters over the 400 iterations
in the left panel of Figure \ref{fig:ClusterHist}. The mode is 17 clusters, which is a substantial
 reduction from the original $L=188$ time series. Figure \ref{fig:ClusterMap} presents a
 representative cluster assignment along with the posterior rates for this
 assignment. This cluster assignment is selected as the assignment that has the minimum
 average Hamming distance across the different iterations \citep[see][for further
 details.]{fox_sticky_2011}.  An interesting phenomenon is that census tracts assigned to
 the same cluster are frequently spatially separated.

\begin{figure}
\centering
\includegraphics[scale=0.7,width=4.75in]{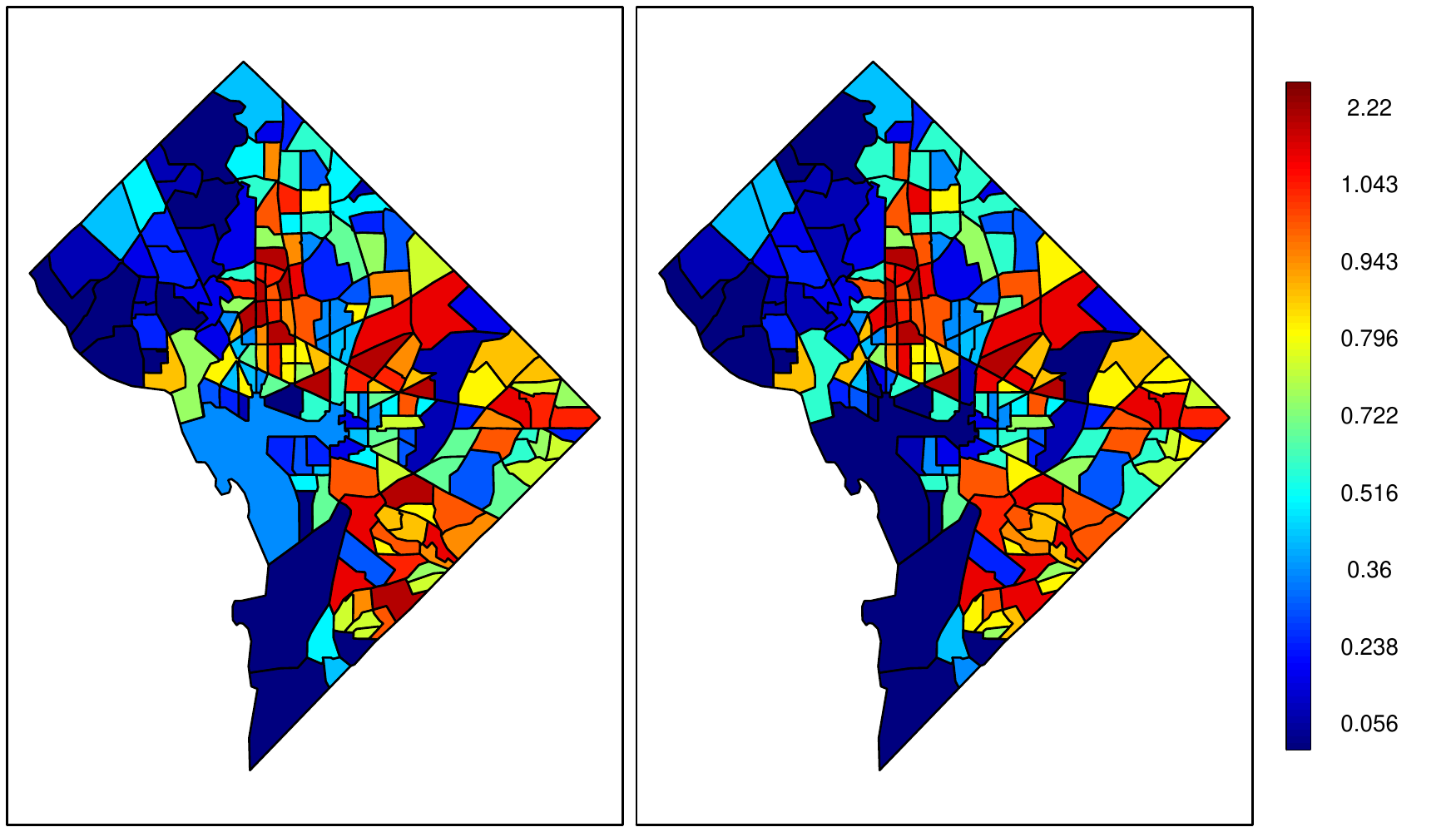}
\caption[Posterior mean rate distribution across the Washington, D.C. map.]{Map of posterior mean rates, $\tractRate_l$, sampled from the multivariate dependent PoINAR(1) model described in Section \ref{subsection:DPoinar} (left) and the population adjusted multivariate dependent PoINAR(1) model described in Section \ref{subsection: cov DoPoinar} (right).}
\label{fig:RateMap}
\end{figure}

\begin{figure}
\centering
\includegraphics[scale=0.8,width=5in, trim = 10 10 10 10]{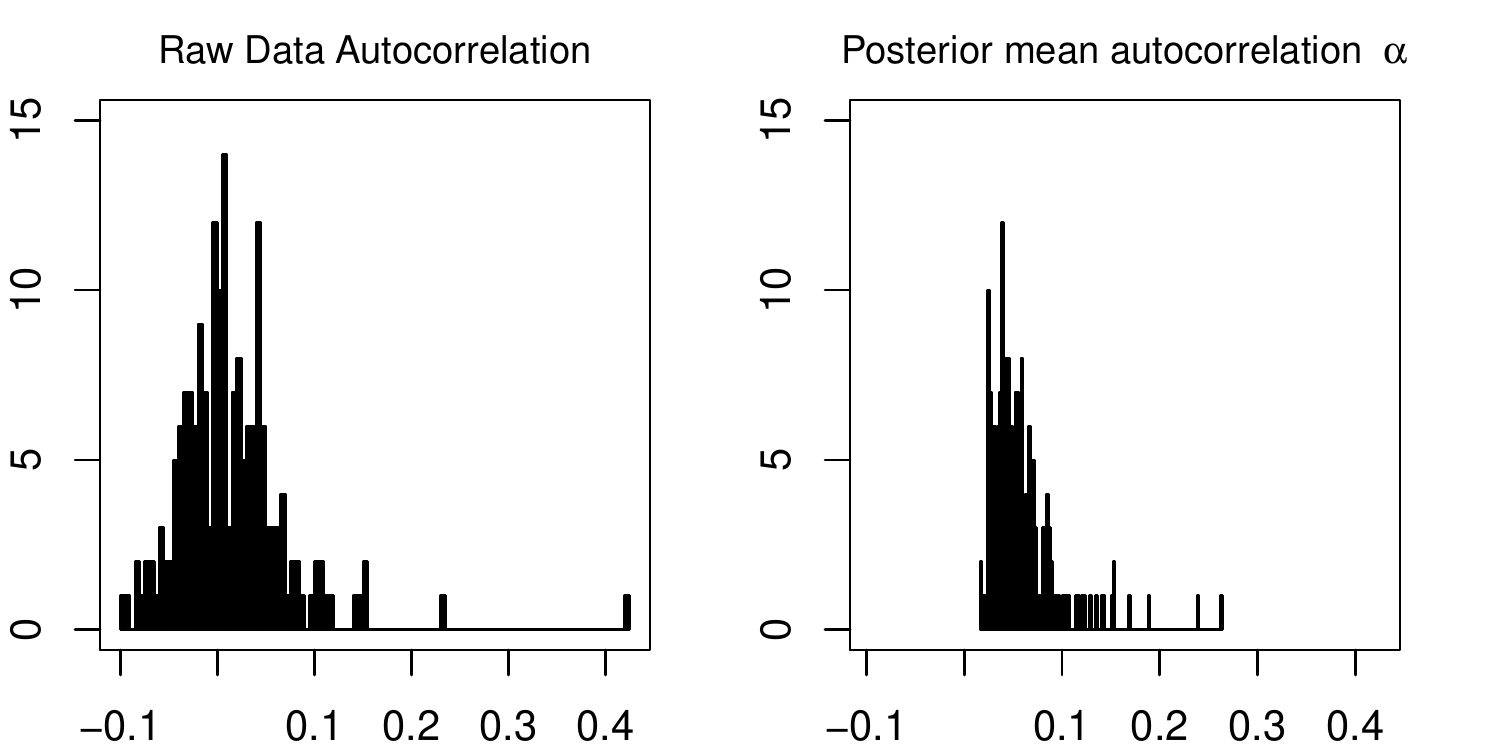}
\caption[Histogram of raw data autocorrelations (left) and posterior mean autocorrelations $\alpha_l$ (right).]{Histogram of raw data autocorrelations (left)
and posterior mean autocorrelations $\alpha_l$ (right).}
\label{fig:HistCorr}
\end{figure}

 We further examine the posterior means for the rates, $\tractRate_l$, of the 188 census
 tracts and their corresponding thinning values, $\alpha_l$, across the MCMC samples.
  Figure \ref{fig:RateMap} (left) maps the posterior mean rates for the census tracts in
 Washington, D.C..  We can see certain regions that exhibit higher rates (e.g., tracts that correspond to a southern portion of the city, a central portion along 16th Street, and an east-central portion along Rhode Island Avenue.)  The results of Figure~\ref{fig:RateMap} are also substantiated
 by Figure~\ref{fig:DC} (right).
 Figure \ref{fig:HistCorr} compares, for each census tract, the sample autocorrelation in
 the counts for each tract with the posterior mean thinning values.  The sample
 autocorrelation is calculated using the classical first order autocorrelation estimator
 for each time series separately (without adjusting for seasonality).  As previously
 explained, the thinning values in our model determine the autocorrelation for each
 INAR(1) time series.  The comparison shows that the raw data autocorrelations vary on a
 wider range of values than their corresponding posterior mean values and some of these
 raw autocorrelations are slightly negative. There are two reasons that can account for
 the differences between the two:
\begin{enumerate}
\item Our model only allows the thinning value to range between $[0,1]$ and therefore
 cannot account for negative autocorrelation.  We believe that the (small) negative raw
 autocorrelations are probably due to noise variation and therefore we are less concerned
 about this phenomenon.  The standard error of an estimated first-order autocorrelation
 for white noise is approximately $1/\sqrt{T} = 1/\sqrt{418} \approx 0.05$; hence the bulk
 of raw autocorrelations are within about 2 standard errors of zero.
\item The posterior mean thinning values are adjusted for seasonal effects, whereas the raw autocorrelations are not. The values would be smaller in magnitude after adjusting for the seasonality, as our results suggest.
\end{enumerate}

 For the out-of-sample evaluation we compare our MCMC method to the CLS and SPP.  For both
 the CLS and our method, we use the estimated one-week-ahead conditional mean as the
 predictor for the crime counts in each tract:
 \begin{align}\label{eq:forecast}
        \hat{\obsVal}_{l,T+1} &= \alpha_l \, \obsVal_{l,T} + \tractRate_l \, \theta_{s(T+1)}.
    \end{align}
 For CLS, we simply plug-in the estimates of $\alpha_l$, $\tractRate_l$ and
 $\theta_m$ for each tract.  For our method, we compute an MCMC-based estimate by
 evaluating Eq.~\eqref{eq:forecast} for each of the 400 MCMC iterations and use the
 average of these as the final predicted value (see Section~\ref{app:simresults} of the
 Supplementary Material for further details). For the SPP, we average
 the past values as the predictor.  

 We predict the one-week-ahead number of crimes in each tract for the first week of each
 month during 2008.  Table \ref{table:ONEWEEKAHEAD} shows the one-week-ahead predicted
 mean RMSE and corresponding standard errors, conditional on the last observed value. The
 results indicate that when the last observed count is one of the most frequent values
 (0,1,2), our method produces lower RMSE.  For the less frequent, higher counts (3,4), the
 performances of all of the methods are (statistically) equivalent.  This behavior is to
 be expected since our method shrinks the estimators toward the mean and therefore should
 perform better for lower, more frequent counts and worse in the more rare cases of high
 counts.  A summary of the average one-week-ahead bias is presented in Section~\ref{app:DCbias} of the
 Supplementary Material. In general, our method produces the smallest bias, but the
 differences between the methods are not significant except when the last observed count
 is zero.

\begin{figure}
\centering
\includegraphics[scale=0.6, trim = 10 10 10 10]{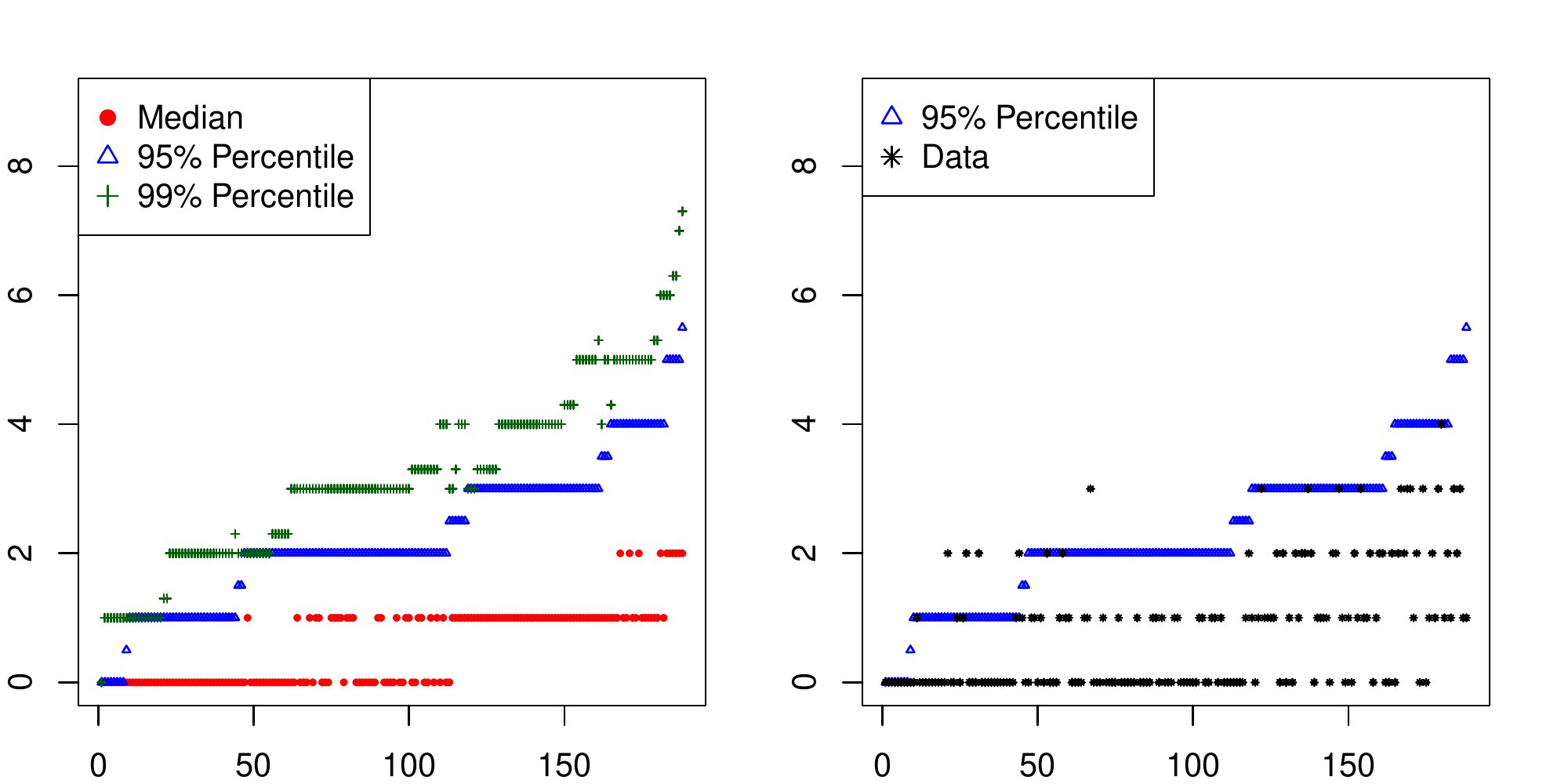}
\caption[The predictive posterior distribution for each of the 188 tracts.]{
The predictive posterior distribution for each of the 188 tracts.
The red dots correspond to the median predicted number of violent crimes for each tract.
The blue triangles and green crosses correspond to the $95\%$ and $99\%$ percentiles of
the predictive posterior distribution, respectively. The black stars corresponds to the test-set actual observed value of crimes.}
\label{fig:predictionInterval}
\end{figure}

 The one-step-ahead conditional mean value is the best linear unbiased estimator under
 quadratic loss. Since the CLS method minimizes the observed squared error, it is only
 natural to evaluate all three methods using the same loss function.  Alternatively,
 \citet{berk_forecasting_2008} proposed a quantile loss function that reflects the
 sensitivity of the police department to forecasting errors.  Under a $\upsilon$-quantile loss function, the predictor is just the predictive
 distribution's $\upsilon$ quantile. Using our method, one can easily sample
 from the following one-step-ahead predictive posterior distribution and evaluate any
 desired quantile:
\begin{multline}
P(\obsVar_{l,t+1}=\obsVal_{l,t+1}|\obsVar_{l,t}=\obsVal_{l,t},\alpha_l^{(m)},\bm \theta^{(m)},\tractRate_l^{(m)}) = \\
\sum_{r=0}^{\infty} {\obsVal_{l,t-1} \choose \obsVal_{l,t}-r}
(\alpha_l^{(m)})^{\obsVal_{l,t}-r} \,
(1-\alpha_l^{(m)})^{\obsVal_{l,t-1}-(\obsVal_{l,t}-r)}\frac{e^{-\uniqueRate_l^{(m)} \,
\theta^{(m)}_{s(t+1)}} \, (\uniqueRate_l^{(m)} \, \theta^{(m)}_{s(t+1)})^r}{r!} \;,
\end{multline}
where $\tractRate_l^{(m)}=\uniqueRate_{z_l^{(m)}}^{(m)}$, $\bm \theta^{(m)}$ and
 $\alpha_l^{(m)}$ are the rate, seasonal component and thinning value estimated during the
 $m^{\textrm{th}}$ iteration of the MCMC sampler.  Figure \ref{fig:predictionInterval}
 shows the $95\%$ and $99\%$ quantiles for each of the 188 tracts and the corresponding
 one-step-ahead true value, $\obsVal_{l,T+1}$.  The quantiles may also be used to provide
 prediction intervals for each tract.  A police department can use these intervals along
 with the point estimate to distinguish between an unusual surge in crimes which requires
 allocation of more resources, and a random rise in crimes, which would not benefit from
 an intervention.  

\begin{table}[h!]
\begin{centering}\footnotesize
\begin{tabular}{|c|c|c|c|c|c|c|}
\hline
$\obsVal_{\cdot,T}$ & 0 & 1 & 2 & 3 & 4 & Overall \\
\hline
\hline
SPP RMSE & 0.8373 & 0.966 & 1.1829 & 1.4722 & 1.4252 & 0.970\\
& (0.034) & (0.0311) & (0.0453) &  (0.088) & (0.1631) & (0.0167)\\
\hline
CLS RMSE & 0.7729 & 0.9501 & 1.0605 & 1.1370 & 1.3258 & 0.9235 \\
& (0.0245) & (0.0430) & (0.0660) & (0.0982) & (0.1991) & (0.0368) \\
\hline
Dependent PoINAR RMSE & \textbf{0.7222} & \textbf{0.9172} & \textbf{1.009}& \textbf{1.0225} & \textbf{1.1600} & \textbf{0.72168} \\
& (0.0135) & (0.0172) & (0.4336) & (0.0862) & (0.1782) & (0.0016) \\
\hline
Frequency & 0.5900 & 0.2340 & 0.1160 & 0.0400 & 0.0200 & 1\\
\hline
\end{tabular}\caption[One-step-ahead root mean squared error analysis as a function of the last observed value of $\obsVal_{\cdot,T}$.]
{One-step-ahead average RMSE as a function of the last observed value of $\obsVal_{\cdot,T}$. We also provide
the standard errors associated with the average RMSE.}\label{table:ONEWEEKAHEAD}
\end{centering}
\end{table}

\section{Covariates Adjusted Dependent PoINAR(1)}\label{subsection: cov DoPoinar}
 Previous research has shown that the tracts of crime are associated with demographic
 covariates, and we have several ways to incorporate such features into our Bayesian
 model. For example, \citet{blei_distance_2009} incorporate
 covariates directly into the clustering mechanism.  This approach might improve the
 accuracy of forecasts, but it would provide less in the way of interpretation, such as
 how the various covariates associate with crime rates.  Instead, we take a more direct
 approach that offers the advantages of clustering as well as interpretation. We model the
 tract-specific rate $\lambda_l$ as a linear function of covariates and cluster the
 coefficients of the equation.  The clusters of coefficients may provide further insight
 into the relationships between crime and demographic characteristics.
%
\subsection{Adjusting for population}
\label{sec:covadjPoINAR}
The main goal of this section is to demonstrate how to add covariates to our model and to
 explore the benefits of doing so.  To this end, we look at the population sizes in each of
 the census tracts as a possible explanatory variable. 

Let $X_l$ denote the population of the $l^{\textrm{th}}$ census tract. (We obtain the
 populations from the 2000 census. Section~\ref{app:DCpop} of the Supplementary Material shows a map of
 the population density in Washington, D.C..)  To incorporate population into our model, we
 redefine the tract-specific rate as a linear function of population,
 i.e. $\tractRate_l=X_l \, \tractAdjRate_l$, where $\tractAdjRate_l$ is the number of
 violent crimes per person in the $l^{\textrm{th}}$ tract. We then place a DP prior
 directly on the rate per person parameter, $\tractAdjRate_l$, yielding the following
 model:
\begin{align}
\begin{aligned}
\epsilon_{l,t} &\sim \textrm{Poisson}(X_l \, \tractAdjRate_{l} \, \theta_{s(t)}) \quad l=1,\ldots,L \quad t=1,\ldots,T\\
\theta_m & \sim  \textrm{F}(\omega) \quad m=1,\dots,12\\
\tractAdjRate_{l} &\sim \textrm{G} \quad l=1,\dots,L\\
\textrm{G} &\sim  \textrm{DP}(\tau,G_0).
\end{aligned}
\end{align}
It is straightforward to adjust the MCMC sampler described in Section \ref{section:MCMC}
 to incorporate the population covariate, $X_l$.  We change the base measure $G_0$ to
 Gamma(0.5,0.5) to reflect the adjustment for population sizes while remaining weakly
 informative.  After these simple modifications, we run the sampler in the manner previously
 described in Section \ref{section:MCMC}.  
\subsection{Analysis of results}
\begin{figure}
\centering
\includegraphics[scale=0.6,trim= 0 0 80 0]{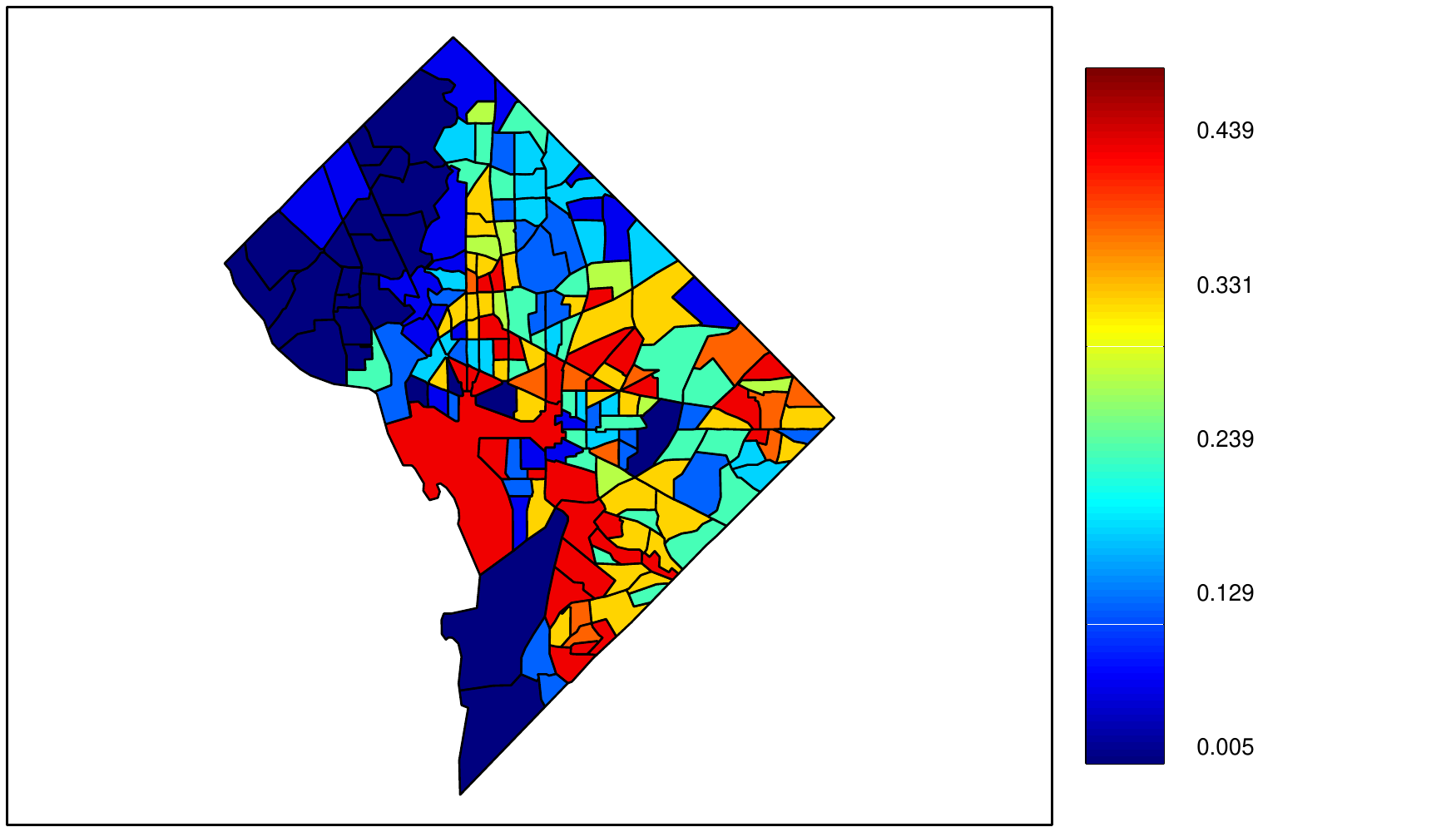}
\caption[Posterior mean rate distribution across the Washington, D.C. map.]{Map of the posterior mean values for crime rates per person, $\tractAdjRate_i$.}
\label{fig:RateMapPop}
\end{figure}

Using the covariate-adjusted PoINAR(1) of Section~\ref{sec:covadjPoINAR}, we again analyze
 the counts of violent crimes in Washington.  As in Section \ref{section:results}, we
 begin by showing the posterior distribution of the number of clusters over the 400 MCMC
 iterations (again taken from 5 chains, each run for 5,000 iterations).  Figure~\ref{fig:ClusterHist} (right) indicates that the distribution is much narrower when we adjust for the population density, and has a mode of 14 clusters.  This suggests that population can account for a significant amount of the spatial heterogeneity in crime. Figure~\ref{fig:RateMapPop} maps the posterior means of the crime rate per person,
 $\tractAdjRate_l$. This map highlights three main features:
\begin{enumerate}
\item The center of Washington, D.C. has a high count of violent crimes per person.
\item The northwest portion of the city has very few crimes per person.
\item The city has three hot-spots: in the center of the city and in the eastern and
 southwestern portions of the city.
\end{enumerate}
These insights, also highlighted by \cite{DC_report}, differ from the conclusions
 one would make by simply looking at the mean values as shown in Figure
 \ref{fig:DC} or from our previous analysis. The results emphasize tracts which exhibit high
 crime rates per person as opposed to high crime counts and can help police make future
 planning decisions, such as where to place a new station.  These outcomes, important as
 they may be, are merely a byproduct of our estimation method.  The more interesting
 research question is whether the population covariate can improve the prediction
 abilities when compared to the unadjusted model.

We performed a one-week-ahead forecast for the last week in the series using both the
 unadjusted model of Section~\ref{subsection:DPoinar} and the adjusted model accounting
 for population.  The overall RMSE of the unadjusted model was 0.9663 whereas the adjusted
 model was 0.9713.  These results suggest that adding the population of the tract to the
 model does not necessarily improve predictive accuracy.  However, a more extensive
 analysis would be needed to confidently settle such issues.  Although adding the
 population of the tract to the model may not improve predictive accuracy, adding the
 covariate seems to provide a useful benefit in that the revised model reveals a more
 interpretable grouping of the time series.  Of course, there are many other covariates
 that one could consider, for example measures of poverty, housing characteristics, etc.
%
%
\section{Discussion}
 In this paper we have presented a method of forecasting multiple correlated low-count time
 series building on the univariate PoINAR framework. The model induces correlation between
 the different time series through two sources: an overall temporal seasonal effect and a
 clustering on individual rate parameters.  The latter clustering is induced by a
 Dirichlet process, which encourages sparse representations in terms of a small number of
 clusters. The grouping of the different rates allows our model to borrow strength across
 the different time series, shrinking the estimators to provide better out-of-sample
 forecasts.

 Our model assumes that there is some underlying clustering assignment of the multiple
 time series.  Moreover, once these clusters are identified, they remain fixed throughout
 time.  One can relax this assumption and allow for temporally evolving cluster
 assignments.  There are a few ways to create such a mechanism, for example we might
 impose dependent Dirichlet process priors, such as those examined by \citet{taddy_autoregressive_2010}.

 Finally, although our focus here is on counts of violent crimes, this model is broadly
 applicable to many low-count spatio-temporal data sets, including the number of insurance
 claims across the U.S., earthquakes across the globe \citep{boudreault_multivariate_2011},
 wildfires across counties \citep{xu_point_2011}, and so forth.

\bibliographystyle{plainnat}
\bibliography{CrimeBib}


\newpage
\setcounter{page}{1}
\begin{center}

\textbf{SUPPLEMENTARY MATERIAL: \\Spatio-temporal Low Count Processes with Application to Violent Crime Events}

\end{center}

\appendix

This supplementary material provides further details on our MCMC sampler in Section~\ref{section:appendix mcmc} and the baseline models to which we compare in Section~\ref{Section:appendix cls}. In Section~\ref{app:sim}, we elaborate upon the simulation studies outlined in the main paper.  Finally, in Sections~\ref{app:DCpop}-~\ref{app:DCbias}, we provide additional details on our Washington, D.C. crime data and results.

\section{MCMC sampler derivation}\label{section:appendix mcmc}
In this section, we detail the derivation of the sampling steps presented in Section~\ref{section:MCMC} of the main paper. We first introduce the following notation:
\begin{itemize}
\item $S_i = \sum_{t=1}^T \epsilon_{i,t}$
\item $\bm S = [S_1,\ldots, S_{N}]$
\item $\bm S_{-i} = [S_1,\ldots, S_{i-1}, S_{i+1},\ldots, S_N]$
\item $z_{-i} = [z_1,z_2,\ldots,z_{i-1},z_{i+1},\ldots,z_N]$
\item $\Theta = \sum_{t=1}^T \theta_{s(t)}$
\end{itemize}

\textbf{\textit{Step 1 -- Sampling the innovations vectors}}
In Step \ref{step:error sample} of the MCMC sampler (Section \ref{section:MCMC}) we motivate the sampling of the innovation vector. We have shown in Eq.~\eqref{eq:sample epsilon} that conditional on the data and other model parameters, the innovations are independent of each other and we only need to sample $\epsilon_{i,t}$ when both the current observed value, $y_{i,t}$, and the previous observed value, $y_{i,t-1}$, have positive values. The values of $\epsilon_{i,t}$ in this case will range between $max\{0,y_{i,t}-y_{i,t-1}\} \leq \epsilon_{i,t} \leq y_{i,t}$. Otherwise, $\epsilon_{i,t}$ is set deterministically.  The distribution of a single innovation, $\epsilon_{i,t}$ is a :
\begin{eqnarray*}
P(\epsilon_{i,t}|y_{i,t-1},y_{i,t},\alpha_i, \lambda^{'}_i,\bm \theta) 
& \propto & P(y_{i,t}|\epsilon_{i,t},y_{i,t-1},\alpha_i)P(\epsilon_{i,t}|\lambda^{'}_i,\bm \theta) \\
\hspace{-0.25in}& = &  {y_{i,t-1} \choose y_{i,t}-\epsilon_{i,t}} \alpha_i^{(y_{i,t}-\epsilon_{i,t})} (1-\alpha_i)^{(y_{i,t-1}-(y_{i,t}-\epsilon_{i,t}))} \frac{e^{-\lambda_{z_i} \theta_{s(t)}}\cdot (\lambda_{z_i}\cdot \theta_{s(t)})^{\epsilon_{i,t}}}{\epsilon_{i,t}!} \\
\hspace{-0.25in} & \propto &  \frac{1}{\epsilon_{i,t}!(y_{i,t}-\epsilon_{i,t})!(y_{i,t-1}-(y_{i,t}-\epsilon_{i,t}))!} 
 \left (\frac{\lambda_{z_i} \theta_{s(t)}  (1-\alpha_i)}{\alpha_i} \right)^{\epsilon_{i,t}} \\
 \hspace{-0.25in}&=&  \frac{1}{C_i}  \frac{1}{\epsilon_{i,t}!(y_{i,t}-\epsilon_{i,t})!(y_{i,t-1}-(y_{i,t}-\epsilon_{i,t}))!}
 \left (\frac{\lambda_{z_i} \theta_{s(t)} (1-\alpha_i)}{\alpha_i} \right)^{\epsilon_{i,t}}
\end{eqnarray*}
We can calculate the normalization constant $C_i$ by summing over all the possible values of $\epsilon_{i,t}$ for a given set of values of $y_{i,t}$ and $y_{i,t-1}$.

\vskip 1cm

\textbf{\textit{Step 2 -- Sampling the membership indicator}}
In the following equations we construct the posterior distribution for the membership indicator variable:
\begin{eqnarray*}
P( z_{i} = j |\bm z_{-i},\bm S ,\bm \theta)
&\propto& P( S_{i}| S_l \quad l\in \{v: z_v=j,v\neq i\},\bm \theta) \cdot P(z_{i}=j |z_l \quad l\in \{v: z_v=j,v\neq i\}).
\end{eqnarray*}
It is straightforward to show that the above distribution has the following form:
\[  P( z_{i} = k |\bm z_{-i},\bm S,\bm \theta ) \propto  \left\{ \begin{array}{ll}
         \alpha \cdot p_{i,0} & \textrm{for } k=K+1 \\
         n_k \cdot p_{i,k} \quad & \textrm{for }k=1,\ldots,K, \end{array} \right. \]
where $p_{i,0},p_{i,1},\ldots,p_{i,K}$ take on values from the following negative binomial distributions:
\begin{eqnarray*}
p_{i,0} &=& \frac{\Gamma(S_i +\gamma_1)}{\Gamma(\gamma_1) S_i!} \left ( \frac{\gamma_2}{\Theta+\gamma_2} \right)^{a} \left ( \frac{\Theta}{ \Theta+\gamma_2} \right)^{S_i} \\
p_{i,j} &=& \frac{\Gamma(S_i+A_j +\gamma_1)}{\Gamma(A_j+\gamma_1) S_i!} \left ( 1- \frac{\Theta }{n_j \cdot \Theta+\gamma_2} \right)^{A_j +\gamma_1} \left ( \frac{\Theta }{n_j \cdot \Theta+\gamma_2} \right)^{S_i} \quad j=1,\ldots,K,
\end{eqnarray*}
and $n_j= \sum_{i=1}^L \mathbb{I}_{z_i=j}$ and $A_j=\sum_{l: z_l=j, l\neq i} S_l$.
This distribution corresponds to the posterior distribution shown in Step \ref{step:membership sample} of the MCMC sampler (Section \ref{section:MCMC}).

\vskip 1cm

\textbf{\textit{Step 3 -- Sampling the unique rate vector}}
Since we use a gamma distribution as the DP base measure, the resulting conditional posterior distribution for the unique cluster-specific rates is as follows:
\begin{eqnarray*}
P(\uniqueRate_k | \bm z, \bm S, \bm \theta, \gamma_1, \gamma_2) &\propto &  P(\bm S_l, \quad l\in \{v: z_v=k\}|\uniqueRate_k, \bm \theta, \gamma_1, \gamma_2)\cdot P(\uniqueRate_k| \gamma_1, \gamma_2) \\
&\propto&  \uniqueRate_k^{B_k+\gamma_1-1} e^{-\uniqueRate_k \cdot (n_k \cdot \Theta +\gamma_2)}
\end{eqnarray*}
This has the form of a gamma distribution with parameters $B_k+\gamma_1$  and $n_k \cdot \Theta +\gamma_2$ where $B_k=\sum_{l\in \{v: z_v=k\}}S_l$
which is the distribution described in Step \ref{step:rate sample} of the MCMC sampler (Section \ref{section:MCMC}).

\vskip 1cm

\textbf{\textit{Step 4 -- Sampling the seasonal effect}}
Let $R_t = \sum_{i=1}^L \epsilon_{i,t}$, then the conditional posterior distribution for the seasonal effect is:
\begin{eqnarray*}
P(\theta_j | \bm \lambda, \bm z, \bm \epsilon, \xi_1, \xi_2) &\propto &  P(\bm \epsilon_{t}, \quad l\in \{t: s(t)=j\}|\bm \lambda, \xi_1, \xi_2)\cdot  P(\theta_j| \xi_1, \xi_2)\\
&\propto &  \theta_j^{ \sum_{t:s(t)=j} R_t + \xi_1-1} e^{-\theta_j \cdot (m_j \cdot \sum_{l=1}^L \lambda^{'}_{l} + \xi_2)}
\end{eqnarray*}
As previously described in Step \ref{step:seasonal sample} of the MCMC sampler (Section \ref{section:MCMC}), this is a gamma distribution with parameters $\sum_{t:s(t)=j} R_t + \xi_1$  and $m_j \cdot \sum_{l=1}^L \lambda^{'}_{l} + \xi_2$ where $m_j= |t:s(t)=j|$.

\vskip 1cm
\textbf{\textit{Step 5 -- Sampling the thinning value}}
The prior distribution for the thinning value $\alpha_l$ is a beta distribution, resulting in the conditional posterior distribution:
\begin{eqnarray*}
P(\alpha_i |\bm y, \bm \epsilon , \eta_1, \eta_2) &\propto & \alpha^{\sum_{t=2}^T (y_{i,t}-\epsilon_{i,t})+\eta_1-1} \cdot (1-\alpha)^{\sum_{t=2}^T (y_{i,t-1}-(y_{i,t}-\epsilon_{i,t}))+\eta_2-1}
\end{eqnarray*}
As previously described in Step \ref{step:thinning sample} of the MCMC sampler (Section \ref{section:MCMC}), this is a beta distribution with parameters $\sum_{t=2}^T y_{i,t}-S_i+\eta_1$ and $\sum_{t=2}^T (y_{i,t-1}-y_{i,t})+S_{i}+\eta_2$.

\vskip 1cm
\textbf{\textit{Step 6 -- Sampling the concentration parametr}}

We follow \citet{escobar_bayesian_1994} and use a gamma distribution prior for the concentration parameter, $\tau$. This stage requires first sampling an
auxiliary variable $\kappa$ which is then used to sample $\tau$:
                \begin{enumerate}
                \item Sample $\kappa\sim\textrm{Beta}(\tau+1,L)$.
                \item Sample $\tau$ as the following mixture of two gammas:
                    \begin{eqnarray*}
                    \tau | \kappa, K &\sim& \pi \textrm{Gamma}(a_{\tau}+K,b_{\tau}-log(\kappa)) \\
                    &+& (1-\pi)\textrm{Gamma}(a_{\tau}+K-1,b_{\tau}-log(\kappa)),
                    \end{eqnarray*}
                    with weight $\pi$ defined by $\pi/(1-\pi)=(a_{\tau}+K-1)/(L \cdot [b_{\tau}-log(\kappa)])$ where $K$ is the number of unique clusters.
                \end{enumerate}

\section{Conditional least squares model}\label{Section:appendix cls}

The PoINAR(1) model can be described in the following manner:
    \begin{eqnarray}
        y_{t+1} &=& \alpha \circ y_t + \epsilon_{t+1} \quad t=1,\ldots,T-1 \\
        \epsilon_{t} & \sim & \textrm{Poiss}(\lambda \cdot \theta_{s(t)})
    \end{eqnarray}
The one-step-ahead conditional expected value for $y_{t+1}$ is:
    \begin{eqnarray}
        \hat{y}_{t+1} &=& \alpha \cdot y_t +\lambda \cdot \theta_{s(t)}
    \end{eqnarray}
The conditional least squares methods estimates this model's parameters by solving the following equation
    \begin{eqnarray}
        \min_{\lambda,\alpha,\theta_1,\ldots, \theta_{12}} \sum_{t=2}^T \left (y_t - \hat{y}_t \right )^2 =
        \min_{\lambda,\alpha,\theta_1,\ldots, \theta_{12}} \sum_{t=2}^T \left (y_t - \alpha \cdot y_t +\lambda \cdot \theta_{s(t)} \right )^2 \quad
        \textrm{s.t.} \sum_{s=1}^{12} \theta_s = 1.
    \end{eqnarray}
This is a nonlinear convex optimization problem. The Lagrangian method yields the following conditions:
\begin{eqnarray}
\alpha &=& \frac{\sum_{t=2}^T y_t \cdot y_{t-1}}{\sum_{t=2}^T y^2_{t-1}} - \lambda \cdot \frac{\sum_{t=2}^T \theta_{s(t)} y_{t-1}}{\sum_{t=2}^T y^2_{t-1}} \\
\lambda &=& \frac{\left ( \sum_{t=2}^T y^2_{t-1}\right)\cdot \left ( \sum_{t=2}^T y_t \cdot \theta_{s(t)} \right) - \left ( \sum_{t=2}^T y_t \cdot y_{t-1}\right) \left ( \sum_{t=2}^T y_{t-1} \theta_{s(t)} \right)}{\left ( \sum_{t=2}^T y^2_{t-1}\right)\cdot \left ( \sum_{t=2}^T \theta^2_{s(t)}\right) - \left ( \sum_{t=2}^T y^2_{t-1} \cdot \theta_{s(t)} \right)^2} \\
\theta_i &=& \frac{2\cdot \lambda \sum_{t:s(t)=i} \left ( y_{t}- \alpha \cdot y_{t-1} \right)- C}{2 \lambda^2 \cdot n_i} \quad i=1,\ldots,12 \\
C &=& \frac{2 \cdot \lambda}{\sum_{i=1}^{12} \frac{1}{n_i}} \left( \sum_{i=1}^{12} \left [ \frac{ \sum_{t:s(t)=i} \left ( y_{t}- \alpha \cdot y_{t-1} \right)}{n_i}\right ] - \lambda\right)
\end{eqnarray}
Starting from a set of initial values, we can iterate between these equations and converge to a solution. The convergence is met
within a few cycles.
\section{Simulation study} \label{app:sim}
To assess the performance of our model, we simulate 9 different datasets from our multivariate PoINAR(1) process.
Each dataset has $L=100$ time series (locations) with $T=208$ observations. The multiple time series are grouped into four equally sized clusters defined by a shared rate value, $\uniqueRate_k$. The different data sets vary in the levels of separation between the cluster rates and the time series autocorrelation values, $\alpha_l$.
In this section, we evaluate the performance of our methods both in- and out-of-sample. The results show that our model can reasonably recover the ground-truth clusterings and also produce accurate out-of-sample forecasts under various settings. Our model also outperforms the
conditional least-squares model (CLS), which is detailed in Section \ref{Section:appendix cls}. One reasonable explanation for these results is that the CLS model does not allow for sharing of information between the time series and therefore is more prone to noise variation.
\subsection{Simulations settings}
We have two main factors that we configure in each of the simulated data sets:
\begin{itemize}
\item The clusters' assigned rate values $\uniqueRate_k$. We examine an ``easy'' setting in which the four cluster rate values are $1,3,6,10$, a ``medium setting'' with values $0.01,0.5,1.2,2$ and a ``hard'' setting with values $0.1,0.2,0.3,0.6$. The rates values are well separated
in the easy setting and becomes harder to distinguish as we move to the hard setting.
\item The thinning value $\alpha_l$, which directly relates to the autocorrelation values of the individual PoINAR(1) processes. We use
three different thinning values shared between all locations: $0.1,0.5,0.9$.
\end{itemize}

\begin{figure}[t!]
\centering
\includegraphics[scale=0.9]{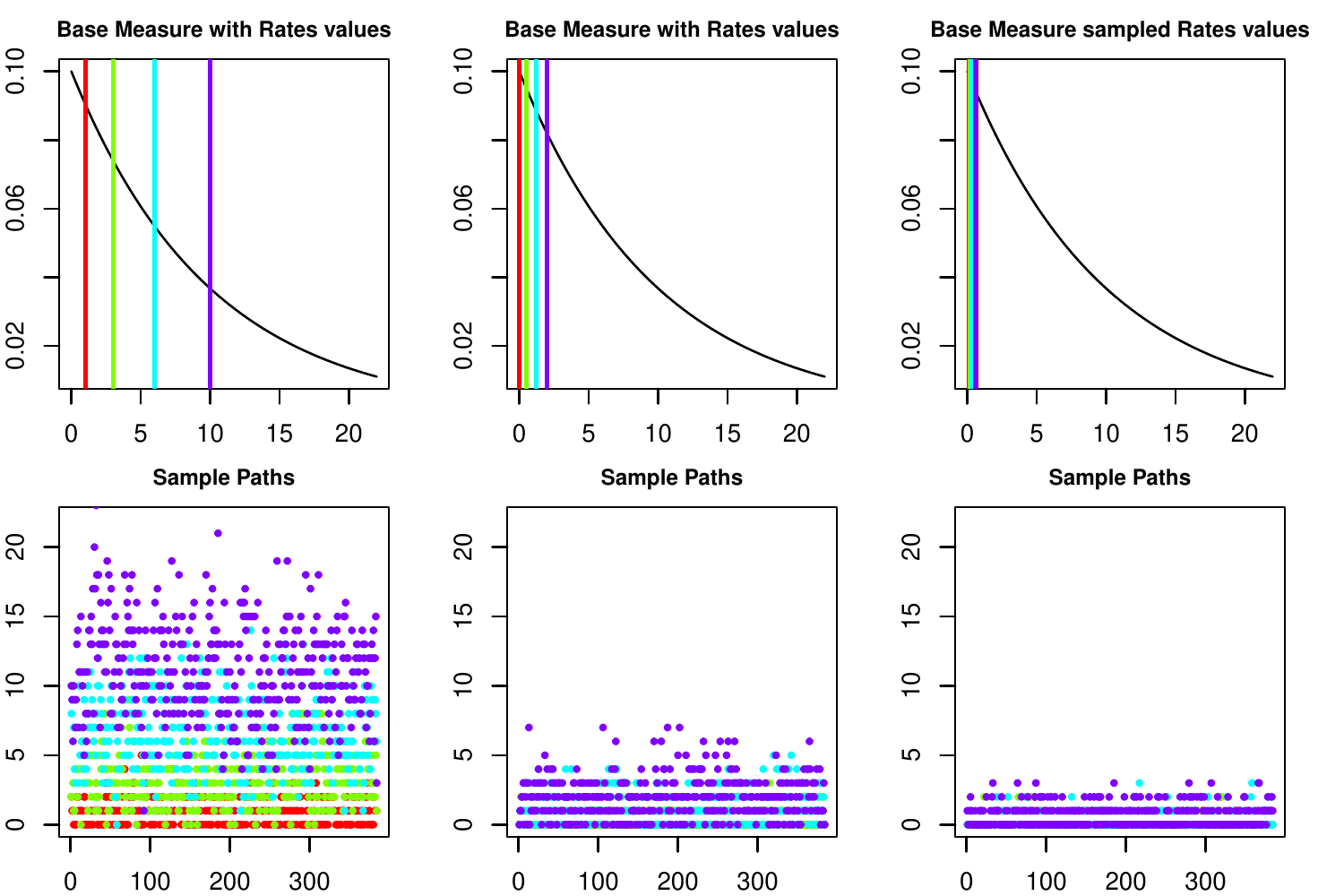}
\caption[Simulations set up.]{The top panel shows the rates values for the four different clusters, along with the prior distribution for the rates.
The lower panel shows an example of 4 overlaid simulated time series. Each time series corresponds to a different cluster and is colored accordingly.}
\label{fig:Sim Set Up}
\end{figure}

Figure \ref{fig:Sim Set Up} illustrates examples of the simulated data for the three different rate scenarios using a thinning value of $0.3$. In the ``easy'' setting, the tract means fall into four clear clusters. In the ``hard'' setting, it is much more difficult to distinguish between the four clusters solely based on the tract means. Furthermore, we can see that as the thinning value grows the tract means become larger and consequently it is easier to identify the clusters.
%
\subsection{Simulation results}\label{app:simresults}
Although we are primarily interested in the out-of-sample performance of our method, there are still two important measures
that are useful to examine in-sample: (i) How many clusters does our method recover? (ii) How close is the recovered clustering assignment to the true assignment? By finding accurate clusterings, our method can borrow information across the multiple time series yielding more accurate out-of-sample predictions.

\begin{figure}[t!]
\centering
\includegraphics[scale=1, trim = 20 10 10 10,width=5.2in]{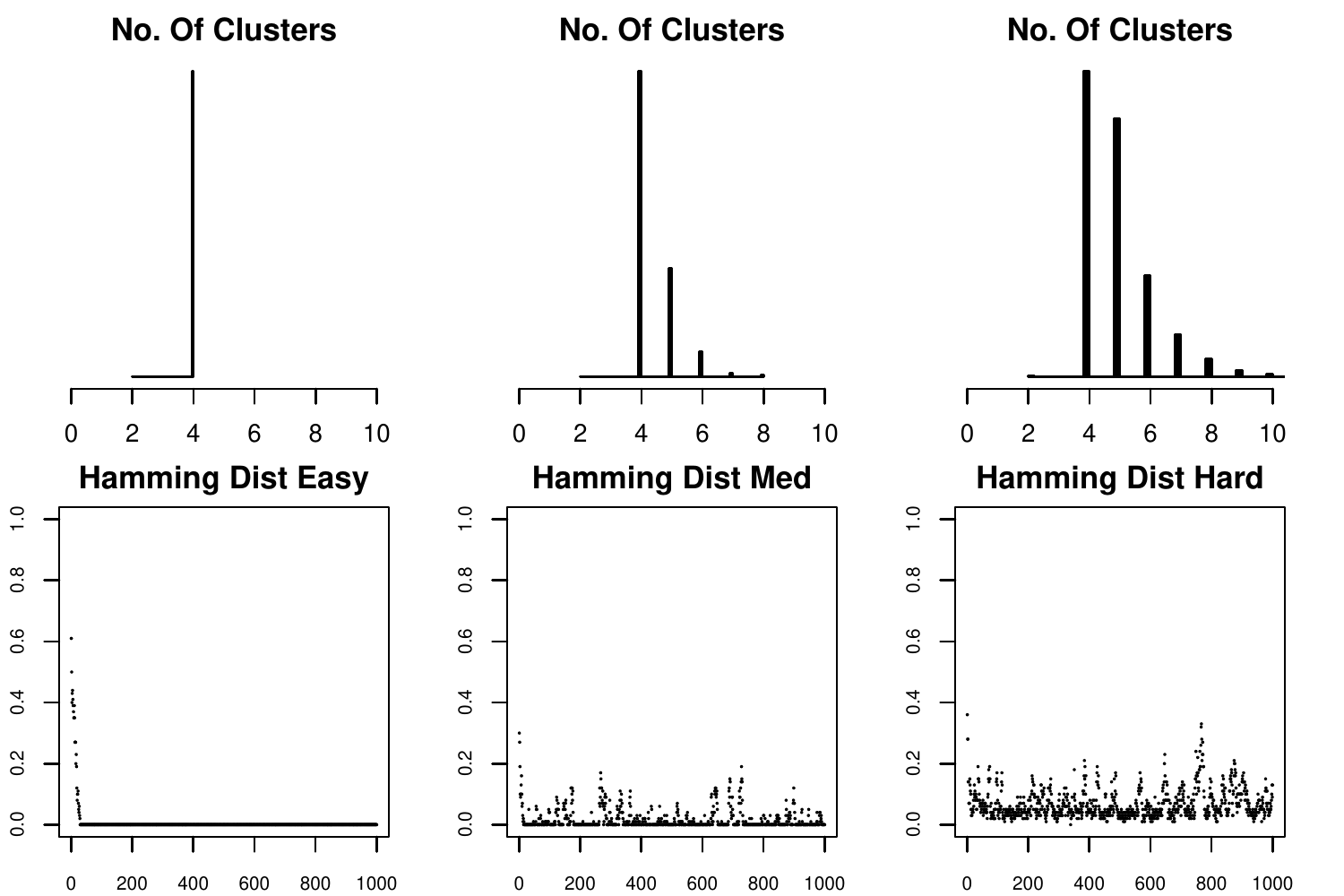}
\caption[In-sample simulations results.]{In-sample simulations results. The top panel displays the histogram for the number of clusters over the $1000$ iterations. The bottom panel shows the Hamming distance errors between the estimated and true cluster assignments versus MCMC iteration.}
\label{fig:Sim Results}
\end{figure}

We run our sampler for 1000 iterations for each of the 9 settings. Figure~\ref{fig:Sim Results} displays histograms for the number of inferred clusters for each of the scenarios plotted in Figure~\ref{fig:Sim Set Up}. Figure~\ref{fig:Sim Results} also shows the Hamming distances between the estimated and true clustering assignment labels. The distances are calculated by first choosing the optimal mapping of indices maximizing overlap between the true and estimated labels assignment sequences. As seen, the modal number of clusters is four in all of the three settings and, as expected, the method recovers the true clustering assignment more accurately for the easy setting than for the hard one. However, even for the hard setting, the Hamming distance errors are usually less than $10\%$ indicating that most time series are correctly clustered. Although we only display the in-sample analysis for these three settings, these results generally hold for all 9 settings.

\begin{figure}[t!]
\centering
\includegraphics[scale=1.2]{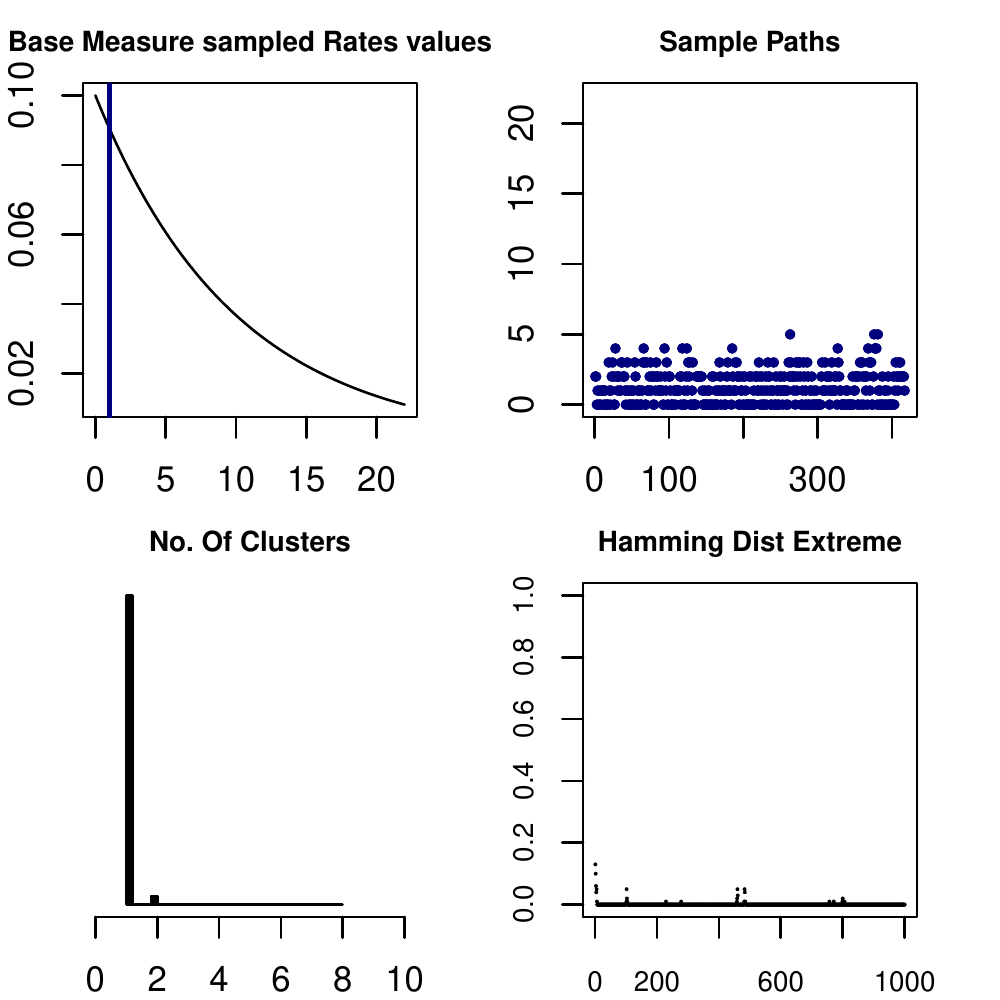}
\caption[Simulation results for the sanity check case.]{Simulation results for a case where there is a single cluster. The top panels display the single cluster rate value and an example of one of the time series' count data. The bottom panels display the histogram of the number of clusters our method finds and the corresponding Hamming distance errors versus MCMC iteration.}
\label{fig:Sim Results sanity}
\end{figure}

An interesting question is whether the methodology finds clustering structure when in fact all the time series belong to the same cluster. To examine this, we simulate a data set that has all of the time series grouped into a single cluster. Figure~\ref{fig:Sim Results sanity} shows the data and the results for the corresponding MCMC sampler. The model predominantly prefers to group all of the time series together, as we would hope in such a case.

In order to evaluate our model's estimation performance, we compare the estimated one-step-ahead conditional expectation of the PoINAR(1) versus its corresponding true (simulated) population value. \citet{braennaes_estimation_1993} showed that the $h$-step-ahead conditional expectation of the PoINAR(1) model is:
    \begin{eqnarray}\label{eq:forecast_app}
        \hat{y}_{T+h} &=& \mathbb{E}(y_{T+h}|y_1,\ldots,y_T) \nonumber\\ 
       \nonumber &=& \mathbb{E}(\alpha^h \circ y_T + \sum_{j=1}^h \alpha^{h-j} \circ \epsilon_{T+j}|y_{T})\\
       &=& \alpha^h \cdot y_T + \lambda \sum_{j=1}^h \alpha^{h-j} \cdot \theta_{s(T+j)}.
    \end{eqnarray}
To use this predictor for our multivariate PoINAR(1) process, we need to produce estimators for the rates value, $\bm \lambda$, the seasonal effects values,
$\bm \theta_s$, and the thinning value, $\bm \alpha$, for the $L$ time series. To this end, we run the MCMC sampler for $m=1000$ iterations and discard the first
$100$ of them as burn-in. We then thin every $5^\textrm{th}$ iteration which leaves us with $180$ iterations from which to infer the parameters in our model. The one-step-ahead conditional expected value for the $m^\textrm{th}$ iteration is:
   \begin{eqnarray}\label{eq:forecastMCMC}
        \hat{y}_{T+1}|\lambda^{'(m)}_i, \alpha^{(m)}_i, \bm \theta^{(m)}  &=& \alpha^{(m)}_i \cdot y_T +\lambda^{'(m)}_i \cdot \theta^{(m)}_{i,s(T+j)}
    \end{eqnarray}
For each time series (location), we now have samples from the posterior distribution of the conditional expected value which we average to produce the corresponding estimated value. We compare the performance of our method with two benchmark methods: the conditional least-squares (CLS) method and a simple Poisson process (SPP). Since the CLS method models the
PoINAR(1) process for each time series separately, we estimate its parameters correspondingly. We then plug these estimators into \refeq{eq:forecast_app} to produce
the corresponding one-step-ahead predicted value for each of the time series. For further details on the CLS method, the reader is referred to Section
\ref{Section:appendix cls}. The SPP assumes, for a single time series, the observed counts are independent identically distributed
Poisson random variables with a constant rate value, $\lambda$. Therefore, we estimate $\lambda$ for each time series using its corresponding counts average and then use this as the one-step-ahead predictor.

To evaluate the different methodologies we use root mean square error (RMSE) and absolute percentage error (APE) between the true population expected value and
its corresponding estimated value based on the $L=100$ time series. The results of this analysis are presented in Table \ref{table:SimRes2app}. The analysis reveals
that our method consistently yields more accurate results compared to the CLS method and the SPP. As expected, the better the separation between
the cluster rate values, the easier it is for our method to estimate the parameters more accurately. In addition, generally higher autocorrelation values produce lower APE but higher RMSE. Intuitively because the stationary distribution mean value for the PoINAR(1) process is
$        \mathbb{E}(y)= \frac{\lambda}{1-\alpha}$,
higher autocorrelation, $\alpha$, yields a higher marginal mean value (or alternatively higher count values). This indicates a larger
separation between the clusters counts values for the data sets with higher $\alpha$. Therefore, higher autocorrelation helps our method identify the ``true'' clusters and yield more accurate estimators based on shrinking.

In conclusion, we believe that because the CLS and SPP methods consider each time series separately, they are more prone to over-fitting.
The proposed Bayesian methodology allows the estimates to pool information from several time series resulting in more robust parameter estimates.
\begin{table}[t!]\footnotesize\centering
\begin{tabular}
{|l||c|c|c||c|c|c||c|c|c|}
\hline
Thin& \multicolumn{3}{c}{0.1} & \multicolumn{3}{|c|}{0.5} &  \multicolumn{3}{c|}{0.9} \\
\hline
Rates & Easy & Med & Hard & Easy & Med & Hard & Easy & Med & Hard \\
\hline
\hline
SPP RMSE &  0.477 & 0.113 & 0.005 & 1.674 & 0.880 & 0.293 & 6.128 & 1.155 & 0.552\\
\hline
CLS RMSE & 0.306 & 0.080 & 0.035 & 0.284 & 0.114 & 0.057 & 0.343 & 0.118 & 0.055 \\
\hline
BNP RMSE & \textbf{0.219} &\textbf{0.058} & \textbf{0.026} & \textbf{0.260} & \textbf{0.086} & \textbf{0.045} & \textbf{0.299} & \textbf{0.075} & \textbf{0.043}  \\
\hline
\hline
SPP APE & 0.067 & 0.159 & 0.1241 & 0.1958 & 0.3192 & 0.2013 & 0.1230 & 0.3372 & 0.2737 \\
\hline
CLS APE & 0.041 & 0.142 & 0.110 & 0.024 & 0.120 & 0.093 & 0.006 & 0.090 & 0.047 \\
\hline
BNP APE & \textbf{0.033} & \textbf{0.041} & \textbf{0.072} & \textbf{0.019} & \textbf{0.033} & \textbf{0.044} & \textbf{0.005} & \textbf{0.046} & \textbf{0.022} \\
\hline
\hline
$E(y_{,T+1})$ & 5.383 & 1.001 & 0.317 & 9.861 & 1.848 & 0.591 & 52.161 & 9.908 & 3.0633 \\
\hline
\end{tabular}
\caption[Estimation performance comparison between the SPP, CLS and MCMC methods]{Conditional mean estimation comparison between the CLS, SPP and our Bayesian nonparametric (BNP) method.
The first four rows show the mean square error (MSE) and the absolute percentage error (APE) between the population (true) one-step-ahead conditional mean and
its corresponding estimated value. The last row shows the average population (true) conditional expected value.}\label{table:SimRes2app}
\end{table}

\section{Washington, D.C. population density map}\label{app:DCpop}
\begin{figure}[h!]
\centering
\includegraphics[scale=0.7, trim = 10 10 10 10]{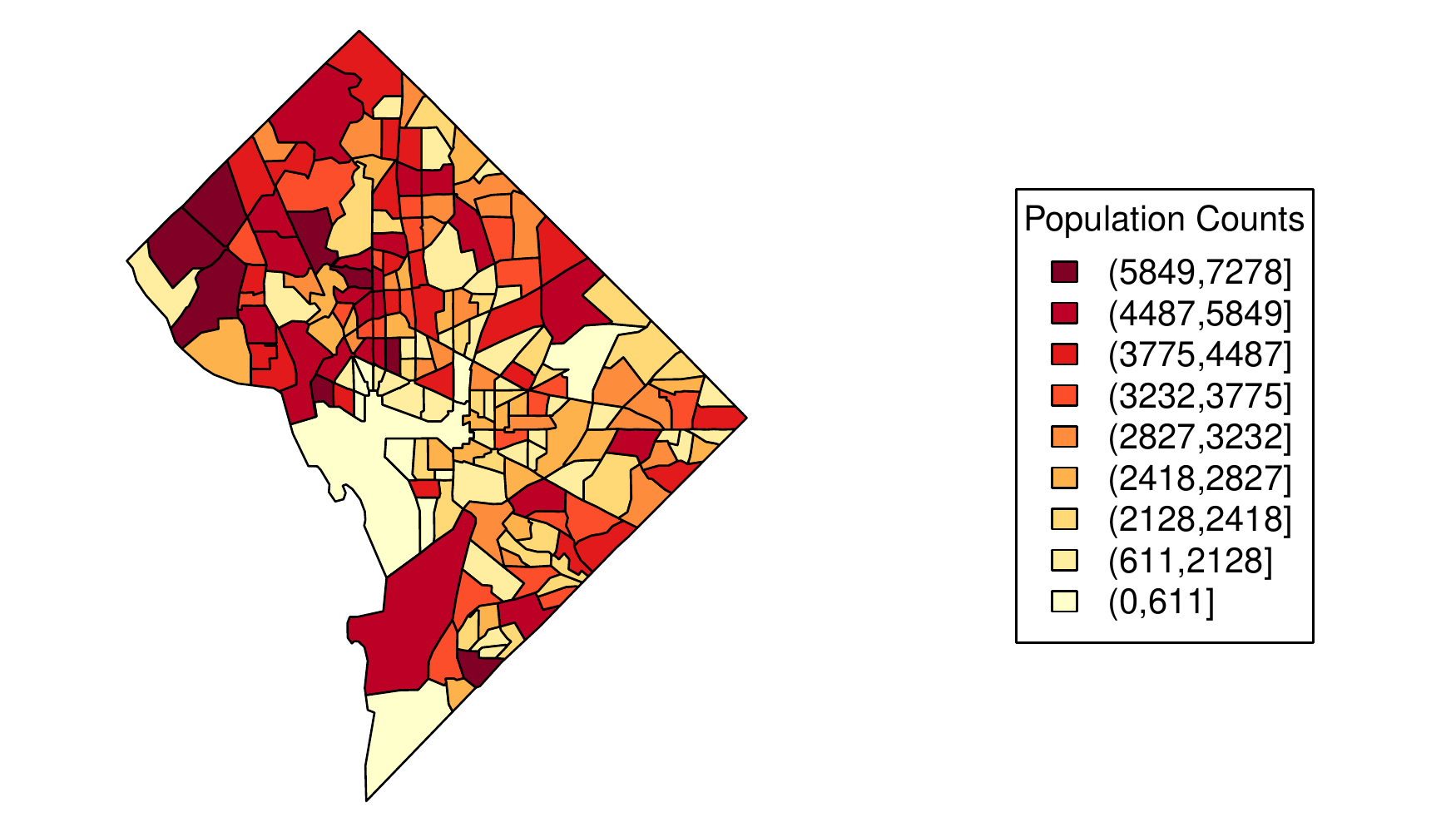}
\caption[Washington D.C. map of population density across the 188 census tracts.]{Washington, D.C. map of population density across the 188 census tracts.}
\label{fig:pop2000}
\end{figure}

Figure \ref{fig:pop2000} shows the map of population density across the 188 Washington, D.C. census tracts.

\section{Washington D.C. crimes time series} \label{app:DCtime}
\begin{figure}[h!]
\centering
\includegraphics[width=6in]{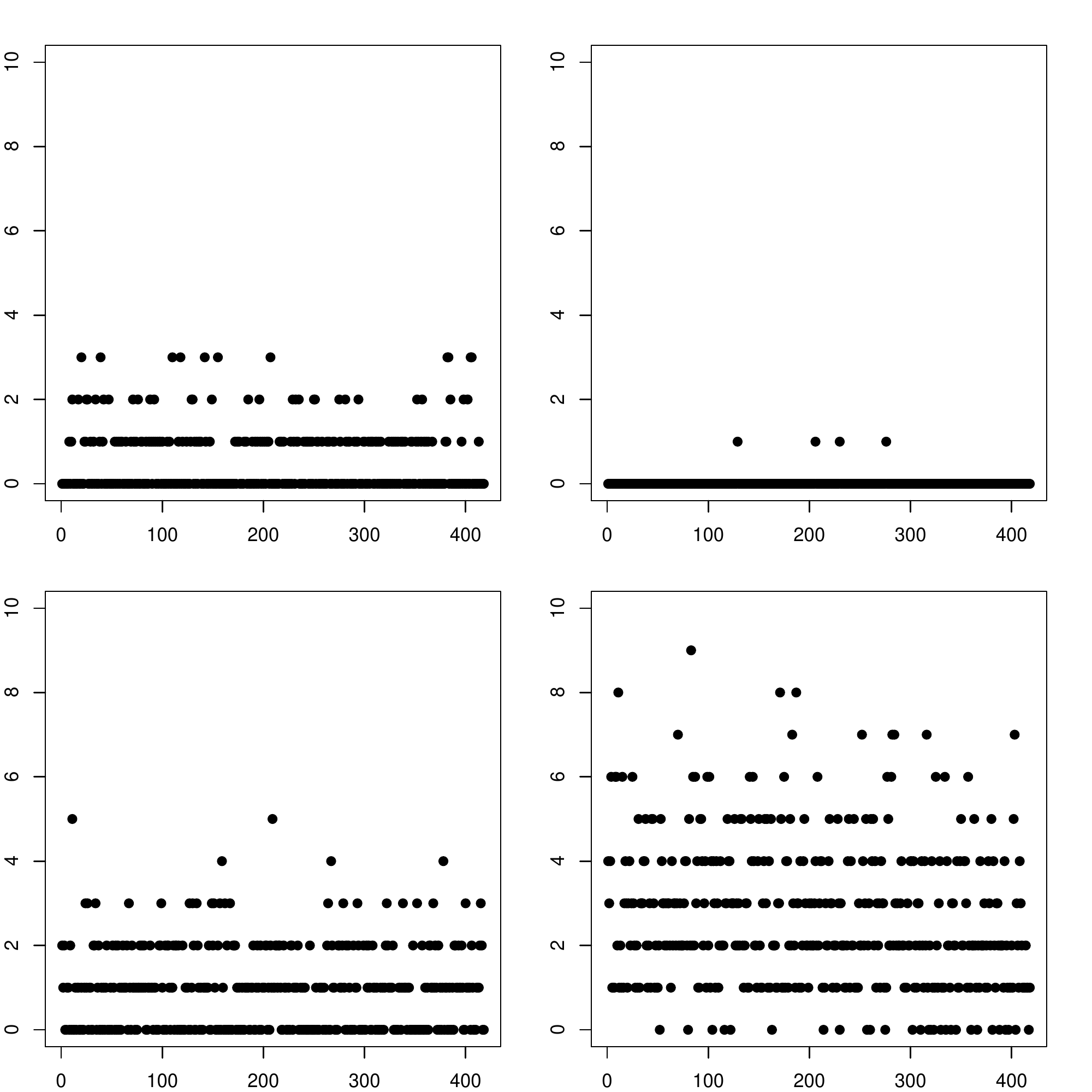}
\caption[Weekly violent crime time series for 4 census tracts.]{Weekly violent crime counts between 2001 and 2008 in 4 census tracts.}
\label{fig:DCts_app}
\end{figure}

Figure \ref{fig:DCts_app} shows the time series of
weekly counts of violent crimes in four tracts between 2001 and 2008. We again see that
some tracts can have very few occurrences whereas others have as many as 9 violent crimes per
week. Also, since the counts are both discrete and small, it is hard to see clear
seasonality within the weekly series.

\section{Bias analysis}\label{app:DCbias}
In Table~\ref{table:ONEWEEKAHEAD_BIAS}, we provide a summary of the average one-week-ahead bias for the Washington, D.C. crime data analysis. As we see from the results, in general, our method produces the smallest bias, but the differences between the methods are not significant except when the last observed count is zero.

\begin{table}[t!]
\begin{centering}
\begin{tabular}{|c|c|c|c|c|c|c|}
\hline
$y_{\cdot,T}$ & 0 & 1 & 2 & 3 & 4 & Overall \\
\hline
\hline
SPP bias & 0.0873 & 0.092& 0.2080 & 0.4730 & 0.5308 & 0.126\\
& (0.0303) & (0.0458) & (0.095) &  (0.1761) & (0.2434) & (0.0379) \\
\hline
CLS bias & -0.1625 & -0.0834 & \textbf{0.1310} & 0.3287 & 0.3849 & -0.073 \\
& (0.0383) & (0.0441) & (0.0716) & (0.1276) & (0.3736) & (0.0405)\\
\hline
BNP bias & \textbf{-0.0234} & \textbf{0.0690} & 0.2175 & \textbf{0.2196} & \textbf{0.19262} & \textbf{0.0456} \\
& (0.0156) & (0.0393) & (0.0710) & (0.1143) & (0.3528) & (0.0232) \\
\hline
Frequency & 0.5900 & 0.2340 & 0.1160 & 0.0400 & 0.0200 & 1\\
\hline
\end{tabular}\caption[One-step-ahead average bias analysis as a function of the last observed value of $y_{\cdot,T}$.]
{One-step-ahead average bias as a function of the last observed value of $y_{\cdot,T}$.}\label{table:ONEWEEKAHEAD_BIAS}
\end{centering}
\end{table}

\end{document}